\documentclass[12pt]{article}
\usepackage{mathrsfs,amssymb,amstext,amsmath,amsthm,eucal}
\usepackage{graphicx,subfig}
\usepackage{longtable}
\usepackage{float}
\usepackage{wrapfig}
\usepackage{rotating}
\usepackage[normalem]{ulem}
\usepackage{marvosym}
\usepackage{slashed}
\usepackage{geometry}
\usepackage{authblk}
\usepackage{tikz}\usetikzlibrary{arrows,shapes}
\usepackage{booktabs,multirow,array}
\newcolumntype{x}[1]{>{\centering\arraybackslash\hspace{0pt}}p{#1}}
\usepackage{enumerate} 

\newcommand{\grad}{\mathrm{d}}

\newcommand{\RR}{\mathbb{R}}
\newcommand{\ZZ}{\mathbb{Z}}

\newcommand{\Spin}{\mathrm{Spin}}

\newcommand{\dvol}{\mathrm{dvol}} 
\newcommand{\lie}{\mathcal{L}}
\renewcommand{\Re}{\operatorname{Re}}
\renewcommand{\Im}{\operatorname{Im}}
\newcommand{\AdS}{\mathrm{AdS}}
\usepackage{hyperref}

\author{Nihat Sadik Deger\footnote{sadik.deger@boun.edu.tr}}
\author{George Moutsopoulos\footnote{gmoutso@gmail.com}}
\affil{Department of Mathematics, Bogazici University, Bebek, 34342, Istanbul, Turkey}
\affil{Feza Gursey Center for Physics and Mathematics, Bogazici University, Kandilli, 34684, Istanbul, Turkey}
\date{\today}
\title{Supersymmetric solutions of $N=(2,0)$ Topologically Massive Supergravity}
\hypersetup{
  pdfkeywords={},
  pdfsubject={},
  pdfcreator={Emacs 24.3.1 (Org mode 8.2.7b)}}

\begin{document}

\maketitle

\begin{center} {\bf Abstract } \end{center}

\begin{quotation}\noindent

We first make a Killing spinor analysis for a general three-dimensional off-shell $N=(2,0)$ supergravity and find conditions for a bosonic background to preserve at least one real supercharge. We then consider a particular model, namely $N=(2,0)$ topologically massive supergravity and impose its field equations. By making a suitable ansatz on metric functions we find a large class of solutions that include spacelike, timelike and null warped $AdS_3$ among others. Isometric quotients of spacelike and timelike squashed $AdS_3$ solutions yield extremal black holes without closed causal curves.
\end{quotation}

\newpage

\tableofcontents
\section{Introduction}
In recent years interest in three-dimensional supergravity theories that admit AdS vacuum increased considerably since they provide laboratories to test various ideas on quantum gravity by using the AdS/CFT duality. 
Off-shell supergravities are good candidates to address these
difficult issues. In these theories there is a lot of freedom in the sense
that one may construct supersymmetric invariant pieces with different number
of derivatives and consider additions of them with arbitrary coefficients.
This big parameter space results in a rich vacua and particular combinations
may have certain advantages such as absence of ghosts.

The isometry group $SO(2,2)$ of $AdS_3$ can be decomposed as $SO(2,1) \times SO(2,1)$ which allows for 
$N=(p,q)$ supergravities \cite{Achucarro:1987vz} in three dimensions where as usual both on-shell and off-shell formulations are possible. 
The $N=1$ \cite{Uematsu:1984zy} and $N=2$ \cite{Rocek:1985bk, Nishino:1991sr, Howe:1995zm, Cecotti:2010dg} off-shell supergravities up to second order derivatives 
has been known for a long time. However, their higher derivative invariants up to fourth order were found 
quite recently in \cite{Bergshoeff:2010mf} and \cite{alkac_massive_2015}, respectively.  
This problem has also been studied using superspace formalism \cite{Kuzenko:2011xg, kuzenko_three-dimensional_2011,kuzenko_three-dimensional_2014,Kuzenko:2015jda}.
The on-shell construction of $N=2$ supergravities were examined in \cite{izquierdo_supersymmetric_1995, Deger:1999st, AbouZeid:2001tu}.

To classify supersymmetric solutions of off-shell models, Killing spinor analysis first developed by Tod \cite{tod_all_1983} is very convenient
since algebraic and differential identities that one obtains remain valid for any combination of higher derivative invariants. This analysis was carried out 
for the $N=(1,0)$ model in \cite{Andringa:2009yc, Bergshoeff:2010mf} and explicit solutions, which are pp-waves, were obtained by assuming that 
the only auxiliary field of the model, i.e. a real scalar, is constant. The off-shell $N=(1,1)$ topologically massive supergravity (TMG) 
contains in addition to Einstein-Hilbert term, the on-shell TMG \cite{Deser:1982sw}, an auxiliary complex scalar and an auxiliary real vector field. 
The supersymmetric solutions of this model without the auxiliary fields were studied in \cite{gibbons_general_2008}.
The Killing spinor analysis of the off-shell $N=(1,1)$ was done in  \cite{deger_supersymmetric_2013} and 
then applied to the off-shell $N=(1,1)$ TMG where class of explicit solutions were found by assuming that the auxiliary vector field is 
constant in a flat basis. When the Killing vector that is constructed from the Killing spinor is timelike these solutions turned out to be 
timelike, spacelike or null warped AdS backgrounds. Later in \cite{alkac_supersymmetric_2015} implications of this Killing spinor analysis and the 
same ansatz was applied to another $N=(1,1)$ model that includes higher derivative invariants of \cite{alkac_massive_2015}. It was found that 
warped AdS solutions of \cite{deger_supersymmetric_2013} are still valid with shifted parameters and two additional solutions, namely 
$AdS_2 \times \mathbb{R}$ and Lifschitz spacetime appear. 

In this paper our goal is to repeat the $N=(1,1)$ analysis of \cite{deger_supersymmetric_2013} for the $N=(2,0)$ case. In the next section we begin with 
reviewing the off-shell $N=(2,0$) multiplet. In section \S\ref{susyconditions} we study implications of the existence of a Killing spinor for a general (2,0)-model. 
As usual the vector that one constructs using Killing spinors turns out to be a Killing vector which is either null or timelike. Since we are dealing with
an off-shell model the results of this section are valid for any (2,0)-theory. This analysis was also carried out in \cite{knodel_rigid_2015} and overlaps
with us. Additionally, in \cite{knodel_rigid_2015} conditions for maximally supersymmetric solutions which admit four supercharges were found and such backgrounds 
were listed. Naturally, not all of them survive once the field equations are imposed as we will see. In section \S\ref{sec:minimal} we introduce the specific model 
that we will apply results of section \S\ref{susyconditions}, i.e. the 
minimal $(2,0)$ TMG, which includes (2,0) off-shell invariants up to third order derivatives. The null case immediately reduces to the on-shell $N=(1,0)$ TMG 
whose supersymmetric solutions are given in \cite{gibbons_general_2008}. In the timelike case by making a suitable ansatz on metric functions we find a 
large number of solutions which we summarize in a table \ref{table}. The assumption we make to find these solutions is weaker than the one made in the $N=(1,1)$ 
case \cite{deger_supersymmetric_2013}, that is, in some of our solutions the vector field is not constant in a flat basis. In addition to warped AdS solutions we 
find several other backgrounds which we summarize in Table \ref{table}.
\begin{table}
\begin{center}
\begin{tabular}{lcccc}
\toprule
Solution & Conditions & Equation 
& $|V|^2$ & Susy
\\\midrule
AdS &  $M\neq 0$ & \eqref{eq:roundads} 
& -
&full
\\
pp-waves & none & \eqref{eq:hnull} 
&  -&
half\\
Minkowski & $M=0$ & \eqref{eq:eucwarp} 
&  - &full
\\
Timelike warped flat& $M(\mu-M/4)=0$ & \eqref{eq:eucwarp} 
&  $<0$&full\\
Timelike warped AdS & $M(\mu-M/4)<0$ \& $\mu\neq -M/4$ & \eqref{eq:wadst} 
& $<0$&half \\
$\RR_t\times H_2$ & $\mu=-M/4$ & \eqref{eq:wadstdegenerates} 
& $<0$&half \\
Lorentzian sphere & $M(\mu-M/4)>0$ & \eqref{eq:lhopf} 
& $<0$&half\\
$\Gamma$-metric & $\mu\neq M/4$ & \eqref{eq:gammametric} 
& any &half \\
$z$-warped null AdS & $M\neq0$ \& $ \mu\neq-M/4$ & \eqref{eq:zwarped} 
&  $0$&half\\
$z$-warped null flat & $M=0$ & \eqref{eq:zflat} 
&  $0$&full \\
Spacelike squashed AdS & $\mu=M/4$ & \eqref{eq:wadss}  
&  $>0$&full\\
Timelike stretched AdS & $\mu=M/4$ & \eqref{eq:wadst2} 
&  $<0$&full\\
Null warped AdS& $\mu=M/4$ & \eqref{eq:wadsn}
&  $0$&full\\\bottomrule
\end{tabular}
\end{center}
\caption{Supersymmetric solutions of (2,0) TMG. We always assume $\mu\neq0$. $|V|^2$ is the typical vector norm, e.g. of $\ast G$ or $V$ in some gauge, and a dash ``-'' means the vectors are identically zero.}
\label{table}
\end{table}
In section \S\ref{sec:asquotients} we consider quotients of warped AdS solution that we have found and using results of \cite{anninos_warped_2009} 
we see that our spacelike and timelike squashed $AdS_3$ solutions yield extremal black hole solutions without closed causal curves.
We finish with a discussion where we also indicate some future directions. In two appendices we give details 
of certain facts that are left out from the main text.

\section{Off-shell $N=(2,0$) Multiplet}
\label{multiplet}
The off-shell three-dimensional  $N=(2,0)$ supergravity multiplet consists of a dreibein $e^a_{\mu}$, a complex gravitino $\psi_\mu$, two gauge 
fields $V_{\mu}$ and $C_{\mu}$, and a real scalar $D$. The supersymmetry transformations\footnote{Our conventions are $\{\gamma_a,\gamma_b\}=+2\eta_{ab}$ 
for a mostly plus signature metric and $\gamma_{\underline{0}\underline{1}\underline{2}}=\epsilon_{\underline{0}\underline{1}\underline{2}}$ in flat 
coordinates $a,b=\underline{0},\underline{1},\underline{2}$. Then, $p$-forms act on spinors via their image in the Clifford algebra, e.g. 
$G\epsilon=\frac{1}{2}G_{ab}\gamma^{ab}\epsilon$ and $K\epsilon=K_{\mu}\gamma^{\mu}\epsilon$. We have defined for convenience the spinor inner product 
such that it is $i$ times that of \cite{alkac_massive_2015}: in our conventions $\bar{\epsilon}\chi$ is anti-hermitian for commuting spinors. Of course, the 
supersymmetry parameter is Grassmann odd in the above transformations.}  of the (2,0) multiplet are
\cite{alkac_massive_2015, kuzenko_three-dimensional_2011,kuzenko_three-dimensional_2014}:
\begin{align}
\delta e_{\mu}{}^{a}& =  \frac{1}{2}i \,\bar{\epsilon}\,\gamma^{a}\,\psi_{\mu}+\text{h.c.}
 \\
\delta\psi_{\mu}& =  \hat{D}_{\mu}\epsilon
\\
\delta C_{\mu}& =  \frac{1}{4}  \,\bar{\epsilon}\,\psi_{\mu}+ \text{h.c.}
\\
\delta V_{\mu}& =   \bar{\epsilon}\,\gamma^{\nu}\hat{D}_{[\mu}\psi_{\nu]} -\frac{1}{4}  \bar{\epsilon}\,\gamma_{\mu}\gamma^{\nu\rho}\hat{D}_{\nu}\psi_{\rho}
- i \bar{\epsilon}\,\hat{G}\,\psi_\mu - D\bar\epsilon\, \psi_\mu +\text{h.c.}
 \\
\delta D & =  -\frac{1}{8} i\bar{\epsilon}\,\gamma^{\mu\nu}\hat{D}_{\mu}\psi_{\nu} +\text{h.c.}
\end{align}
where the supercovariant derivative is
\begin{equation}
\hat{D}_{\mu}\epsilon \equiv \left(\partial_{\mu}+\frac{1}{4}{\omega}_{\mu}{}^{ab}\,\gamma_{ab}
-i V_{\mu}\right)\epsilon- i \, \gamma_{\mu} \hat{G}\epsilon-\gamma_{\mu}\,D\epsilon ~ \label{eq:D}
\end{equation}
and $\hat{G}$ is the supercovariant field strength of $C_{\mu}$. That is, $\hat{G}\equiv G+\mathcal{O}(\psi^2)$ with $G=\grad C$.

The $N=(2,0)$ multiplet transforms under a $U(1)_R$ transformation
as
\begin{subequations}\label{eq:gaugetrans}
\begin{align}
\psi_{\mu} &\mapsto e^{i\phi} \psi_{\mu}\\
V_{\mu} &\mapsto V_{\mu} + \partial_{\mu} \phi ~.
\end{align}\end{subequations}
The supersymmetry parameter $\epsilon$ transforms like $\psi_{\mu}$, that is  $\epsilon\mapsto e^{i\phi}\epsilon$. The supercovariant derivative $\hat{D}_{\mu}$ in \eqref{eq:D} is thus seen to be $U(1)_R$ covariant. We will first analyze the general implications of a bosonic background such that it preserves some supersymmetry. Subsequently we will turn to a specific model.

\section{Implications of Supersymmetry}
\label{susyconditions}
We consider a bosonic background of a (2,0)-invariant theory, that is some data $(g,V,G,D)$ where $g$ is the metric and we set $\psi_{\mu}=0$. A supersymmetric 
background is such a background that admits a Killing spinor. That is, there is a commuting complex spinor  $\epsilon$ that satisfies 
\begin{equation}\label{eq:KS}
\hat{D}_{\mu}\epsilon = \nabla_{\mu}\epsilon
-i V_{\mu} \epsilon- i \, \gamma_{\mu} {G}\epsilon-\gamma_{\mu}\,D\epsilon =0~.
\end{equation}
Following the seminal work of Tod \cite{tod_all_1983} and the analogous analysis of $(1,1)$ supersymmetry in \cite{deger_supersymmetric_2013}, a lot can be derived from the bosonic Killing spinor bilinears. We accordingly define
\begin{align}
K_\mu &\equiv\bar{\epsilon}\gamma_\mu\epsilon \\
if &\equiv \bar{\epsilon}\epsilon~, \\
L_\mu &\equiv \bar{\epsilon}\gamma_\mu\epsilon^{*}~.
\end{align}
Here $f$ is a real scalar, $K_{\mu}$ is real one-form and $L_{\mu}$ is a complex one-form. Note that because of the antisymmetry of the inner product we have
\begin{equation}
\bar{\epsilon}\epsilon^{*}=0~,
\end{equation}
and due to the fact that
\begin{equation}
\gamma_{\mu\nu} = \epsilon_{\mu\nu}{}^{\rho}\gamma_{\rho}~ ,
\end{equation}
all spinor bilinear two-forms and three-forms are related to $f$, $K_{\mu}$ and $L_{\mu}$. As usual, we will show that $K_\mu$ is Killing. There will be algebraic and differential relations involving $f$ and $L_{\mu}$ that will severely restrict the background.

We begin with some algebraic relations. We contract the Fierz identity for commuting spinors
\begin{equation}\label{eq:fierz}
\epsilon_1 \bar{\epsilon}_2 = \frac12 \left(\bar{\epsilon}_2 \epsilon_1\right) I + 
 \frac12 \left(\bar{\epsilon}_2 \gamma_\mu\epsilon_1 \right)\gamma^\mu
\end{equation}
with $\epsilon$ and/or its complex conjugate and we likewise set $\epsilon_1$ and $\epsilon_2$ to be either $\epsilon$ or its complex conjugate. This produces the following relations
\begin{align}
 K_{\mu} K^{\mu} &= -f^2 \label{eq:quadfierz-beg}\\
\left(L_{\mu} \right)  \left(L^{\mu}\right)^{*} &= +2f^2 \\
L_{\mu} L^{\mu} &= 0
\label{eq:L2zero}\\
 K_{\mu}L^{\mu} &= 0~.\label{eq:quadfierz-end}
\end{align}
If $f=0$, then the Killing spinor will be called null and if $f\neq0$  it will be called timelike. In principle, a Killing spinor may smoothly change type from one patch to another over the spacetime, but this will not enter into our analysis. We will assume a Killing spinor that is either timelike or null in the region where we solve the equations of motion. Accordingly, we will talk about a timelike or a null supersymmetric background. 

The importance of the distinction between timelike and null Killing spinors is already obvious from the quartic Fierz identities \eqref{eq:quadfierz-beg}-\eqref{eq:quadfierz-end}. Indeed, for $f\neq0$ we see that the vectors $K_\mu$, $\Re L_{\mu}$ and $\Im L_{\mu}$ constitute an orthogonal frame
 \begin{equation}\label{eq:ortho}
\begin{pmatrix} K_\mu \\ \Re( L_\mu ) \\ \Im( L_\mu ) \end{pmatrix}
\begin{pmatrix} K^{\mu} &  \Re( L^{\mu} ) & \Im( L^{\mu} ) \end{pmatrix} 
= f^2 \begin{pmatrix} -1 & 0 & 0 \\ 0 & +1 & 0 \\ 0 & 0 & +1 \end{pmatrix}~.
\end{equation}
On the other hand, if $f=0$, then $K_{\mu}$,  $\Re( L_\mu )$ and $ \Im( L_\mu )$ are all null and orthogonal to each other, which in lorentzian signature implies that they are also proportional. 

From \eqref{eq:KS} we derive
\begin{equation}
\nabla_\mu K_\nu = \overline{\nabla_\mu \epsilon}\gamma_\nu \epsilon + \bar{\epsilon} \gamma_\nu \nabla_\mu \epsilon = -2 f G_{\mu\nu} - 2 D \epsilon_{\mu\nu\rho}K^\rho ~. \label{eq:killing0}
\end{equation}
From this it follows that $\nabla_\mu K_\nu$ is antisymmetric, whence $K$ is a Killing vector that is either timelike ($f\neq0$) or null ($f=0$) and we rewrite\footnote{We always identify the Killing vector $K$, or indeed any other vector field, with its metric dual one-form. We define the Hodge dual by $\omega\wedge\ast\omega=|\omega|^2_{g} \dvol_{g}$  for a differential form $\omega$ of length squared $|\omega|_{g}^2$. With our signature and dimension, we have $\ast^2=-1$ and $(\ast \grad \ast \omega)^{\cdots} = (-1)^{|\omega|}\omega^{n\cdots}{}_{;n}$ for a differential form $\omega$ of degree $|\omega|$.} the above equation as
\begin{equation}
\grad K = - 4 f G - 4 D \ast K ~. \label{eq:killing1}
\end{equation}
Similarly, we find the derivative of the scalar $f$ to be
\begin{equation}
\partial_\mu f = 2 G_{\mu\nu}K^\nu~, \label{eq:killing2}
\end{equation}
which we may rewrite as
\begin{equation}
i_K G = - \frac{1}{2}\grad f~.\label{eq:killing2b}
\end{equation}
If $f\neq0$ then \eqref{eq:killing2} can be derived\footnote{Indeed, from the antisymmetry of $\nabla_{\mu}K_{\nu}$ the contraction of the left-hand side of \eqref{eq:killing0} with $K^{\nu}$ becomes 
$f \partial_\nu f$ whereas on the right-hand side only the first term survives 
and gives $-2 f G_{\mu\nu} K^{\mu}$.} from contracting \eqref{eq:killing0} with $K^{\mu}$. 
If $f=0$ then the two equations are independent.

By using the fact that $G$ is closed and \eqref{eq:killing2b}, we may show that $K$ preserves $G$
\begin{equation}\label{eq:lieGzero}
\lie_K G = i_K \grad G + \grad i_K G = 0~.
\end{equation}
Furthermore, we may act on \eqref{eq:killing1} with $\lie_K$ and use $\lie_K G = 0$ in order to show that 
\begin{equation}\label{eq:lieDzero}
\lie_KD=0~.
\end{equation} Indeed, $\lie_K$ annihilates $G$, $K$ and $f$ in \eqref{eq:killing1} and commutes with $\ast$ and $\grad$. In summary, $K$ is a Killing vector that preserves  
the fields $G$ and $D$. 
We will later show that $K$ also preserves $V$ up to a gauge transformation. Finally, from the derivative on $L_{\mu}$ we derive the complex equation
\begin{equation}
\nabla_\mu L_\nu = - 2 i V_\mu L_\nu - 2 D \epsilon_{\mu\nu \rho}L^\rho - i \left( G_{ab}\epsilon^{abc}L_c \right)g_{\mu\nu} + 2i \epsilon_{\mu\nu a}G^{ab}L_b~. \label{eq:killing4} 
\end{equation}
We will analyze this equation separately according to whether $\epsilon$ is timelike or null. 

\subsection{Null Killing Spinor}
By writing the spinor inner product $\bar{\epsilon}\chi$ as the $\Spin(1,2)=\mathrm{SL}(2,\RR)$ invariant determinant of $\epsilon^{*} \otimes \chi$, it is seen that the real part and imaginary part of a null Killing spinor $\epsilon$, for which in particular $\bar{\epsilon}\epsilon = 0$, are linearly dependent. 
Equivalently, a null Killing spinor $\epsilon$ is real up to an overall phase,
\begin{equation}
\epsilon^{*} = e^{2ic} \epsilon~.\label{eq:nullrealspinor1}
\end{equation} 
The nonzero Killing spinor bilinears are the null Killing vector $K$ and the complex vector 
\begin{equation}
L = e^{2ic}K~,\label{eq:nullrealspinor2}
\end{equation}
in agreement with our earlier observation after \eqref{eq:ortho} that $K$ and $L$ should be proportional to each other. 

The differential identities \eqref{eq:killing0}, \eqref{eq:killing2} and \eqref{eq:killing4} still hold upon setting $f=0$. 
Substituting $L=e^{2ic}K$ in \eqref{eq:killing4} and using \eqref{eq:killing2} for $\nabla_{\mu} K_{\nu}$ again, we derive a tensor equation that implies
\begin{align}
\epsilon^{abc} G_{ab}K_{c} &= 0 \label{eq:kGzero2} \\
V + \grad c&=0~. \label{eq:Visdc}
\end{align}
The second equation states that $V$ is pure gauge.  We may thus choose the gauge where $V=0$ and $c=0$, for which \eqref{eq:nullrealspinor1} and \eqref{eq:nullrealspinor2} become 
\begin{equation}
\epsilon=\epsilon^{*} \quad \text{and} \quad K=L~.
\end{equation} 
The metric admits a null Killing vector that satisfies, according to \eqref{eq:killing1} with $f=0$,
\begin{equation}
\grad K = - 4 D \ast K~. \label{eq:beltrami}
\end{equation}


It was shown in \cite{gibbons_general_2008} that given a metric that admits  a null Killing vector $K$ for which  \eqref{eq:beltrami} holds with $D=1/2$, there exist coordinates $(x,u,v)$ for which $K=\partial_v$ and such that the metric takes the form
\begin{equation}
\left.g\right|_{D=1/2}= \grad \rho^2 + 2 e^{2\rho} \grad u \grad v + h(u,\rho) \grad u^2~. \label{eq:adsppwave}
\end{equation}
This metric has been referred to as the AdS pp-wave: the natural generalization of a flat space pp-wave where $\partial_v$ is Killing and null but no longer $\nabla$-parallel. The derivation in  \cite{gibbons_general_2008} can be easily generalized to any $D\neq0$, not necessarily constant, so that instead the metric is given by
\begin{equation}
\left.g\right|_{D\neq0}= \frac1{4 D(u,\rho)^2}\grad \rho^2 + 2 e^{2\rho} \grad u \grad v + h(u,\rho) \grad u^2~.\label{eq:gnullDnotzero}
\end{equation}
We give the proof in appendix \ref{app:gib}, where we note that the sign of $D$ is correlated to the sign of orientation. 
The case of $D=0$ corresponds to that of a flat space pp-wave, which in Brinkmann coordinates is given by
\begin{equation}
\left.g\right|_{D=0}= \grad \rho^2 + 2 \grad u \grad v + h(u,\rho) \grad u^2~,
\label{eq:gnullDzero}
\end{equation}
where  $K=\partial_v$ is the parallel null vector. In both cases \eqref{eq:gnullDnotzero} and \eqref{eq:gnullDzero}, we may solve  \eqref{eq:kGzero2}, \eqref{eq:killing2b} with $f=0$, and $\grad G=0$ for the coordinate form of $G$, arriving at
\begin{equation}\label{eq:Gnullcoo}
G = \tilde{G}(u,\rho) \grad \rho \wedge \grad u~.
\end{equation}
The function $\tilde{G}(u,\rho)$ is however left undetermined.

We may contract the Fierz identity \eqref{eq:fierz} with $\epsilon$ in order to prove that 
\begin{equation}\label{eq:Knullani}
K \, \epsilon = 0 ~.
\end{equation}
The solution \eqref{eq:Gnullcoo} and \eqref{eq:Knullani} then imply
\begin{equation}
G \epsilon = 0~.
\end{equation} 
From this, the Killing spinor equation simplifies to
\begin{equation}
\nabla_\mu \epsilon = D \,\gamma_{\mu} \epsilon~.\label{eq:geomkill}
\end{equation}
Let us remark that without imposing equations of motion, $D$ is not necessarily a constant. This equation is the so-called pseudo-riemannian Killing spinor equation, see for instance~\cite{baum_twistor_2000}. 
In higher dimensions such a real spinor $\epsilon$ is not necessarily null. Furthermore, in euclidean signature the function $D$ in \eqref{eq:geomkill} is known to be necessarily either a real constant or an  imaginary function~\cite{lichnerowicz_spin_1987,rademacher_generalized_1991}. In our case, i.e. three-dimensional lorentzian signature, the following natural question arises: whether the integrability condition of \eqref{eq:geomkill} is automatically satisfied for \eqref{eq:gnullDnotzero}, or 
more specifically for \eqref{eq:gnullDzero}. 
In particular, we ask whether the existence of a null Killing spinor 
may further restrict the bosonic data $(h,D)$. In the case of a flat space pp-wave it is known that the riemannian spin holonomy reduces so that in particular %
it 
admits a $\nabla$-parallel real spinor and there are thus no more restrictions. We will now show that the same is true for $D\neq0$, that is \eqref{eq:geomkill} is integrable and the data $(h,D)$ are left undetermined.

In order to proceed, let us use the orthonormal frame
\begin{align}
\theta^{\underline{\rho}} &= \frac{1}{{2|D|}} \grad \rho \\
\theta^+&= \grad u \\
\theta^-&= \frac{h}{2} \grad u + e^{2\rho}\grad v~.
\end{align}
We choose the sign of orientation $\dvol=\theta^{\underline{\rho}}\wedge \theta^+ \wedge \theta^-$, which corresponds to $D<0$. 
The Killing spinor equation \eqref{eq:geomkill} for a Killing spinor $\epsilon'$ in the above frame is written in the appendix \ref{app:ks}. The identity \eqref{eq:Knullani} in the chosen frame becomes
\begin{equation}
\gamma_- \epsilon = 0~.\label{eq:bpsnull}
\end{equation}
Combined with the projection equation \eqref{eq:bpsnull}, the Killing spinor equation becomes\footnote{Note that the orthonormal frame in \cite{gibbons_general_2008} is boosted by a Lorentz transformation with respect to ours and so is their Killing spinor solution for $D=1/2$.}
\begin{align}
\left(\partial_\rho - 1\right)\epsilon &= 0\\
\partial_u \epsilon &= 0\\
\partial_v \epsilon &= 0~.
\end{align}
Let us note that we would arrive at the same equations had we chosen the opposite orientation, for which $D>0$. These equations are easily solved by $\epsilon=e^{\rho}\epsilon_0$ without further restricting $D$ or $h$, where $\epsilon_0$ is a constant spinor.


We summarize that a null Killing spinor of a (2,0) theory implies $V=0$, the field strength \eqref{eq:Gnullcoo} and the metric \eqref{eq:gnullDnotzero} if $D\neq 0$ or \eqref{eq:gnullDzero} if $D=0$. Conversely, a metric of the form  \eqref{eq:gnullDnotzero} if $D\neq 0$ or \eqref{eq:gnullDzero} if $D=0$ is sufficient for at least locally admitting a real spinor that satisfies \eqref{eq:geomkill} and subject to the projection \eqref{eq:bpsnull}. Furthermore, the sign of $D$ fixes the sign of orientation appropriately.
\subsection{Timelike Killing Spinor}
If $f\neq0$ and thus $K$ is a timelike Killing vector, then the metric can be written in adapted coordinates $(t,x^1,x^2)$ so that $K=\partial_t$ and 
\begin{equation}\label{eq:KKform}
g = - f^2 \left(\grad t + A\right)^2 + e^{2\sigma} \left( \grad x^2 + \grad y^2 \right)~.
\end{equation}
In these coordinates we will use the orthonormal frame
\begin{subequations}\label{eq:orthoframe}
\begin{align}
\theta^{\underline{0}} &= f \left( \grad t + A \right) \\
\theta^{\underline{i}} &= e^{\sigma} \grad x^i~.
\end{align}\end{subequations}
Note that since $\epsilon$ is defined globally up to an overall constant scaling then so is the scalar $f=-i\bar{\epsilon}\epsilon$. The Kaluza-Klein vector $A$ transforms under the Kaluza-Klein local $U(1)$ gauge, $t\mapsto t+\chi(x^i)$, and the 2d diffeomorphisms $x^{i}\mapsto h^i(x^j)$. The scalar $\sigma$ breaks the 2d diffeomorphism symmetry but transforms under two-dimensional conformal transformations as a Liouville field. 
It will be useful to define the scalar $\rho$ via
\begin{equation}
\grad A \equiv \rho \, \theta^{\underline{1}} \wedge \theta^{\underline{2}}~,
\end{equation}
or $\rho=2\partial_{[1}A_{2]} e^{-2\sigma}$. Note then that $\rho$ is invariant under the  Kaluza-Klein symmetry and the conformal symmetry of $\sigma$. 
Finally note that scaling time $t$ by a constant will scale $f$ and $\rho$ appropriately but leave their product $f \rho$ invariant. The latter is a freedom that we will use later.

The orthogonality properties \eqref{eq:L2zero} and \eqref{eq:quadfierz-end} of $K$ and $L$ can be rewritten in the orthonormal basis \eqref{eq:orthoframe} as $L_{\mu}K^{\mu}= f L_{\underline{0}} = 0$ and $L_{\mu}L^{\mu}=-  \left(L_{\underline{0}}\right)^2 + \sum_{i=1}^2  \left(L_{\underline{i}}\right)^2=0$. We thus derive
\begin{align}
L_{\underline{0}}&=0 ~, \label{eq:L0}\\
L_{\underline{2}}&=i L_{\underline{1}}~. \label{eq:L12relation}
\end{align}
We note that $K = g(\partial_t,-) = - f \theta^{\underline{0}}$ yields
\begin{align}
 K_{\underline{0}} &= -f ~,\\ 
K_{\underline{i}}&=0~.
\end{align} 
Finally, from $\Re(L_{\mu})\Re(L^{\mu})=f^2$ and \eqref{eq:L12relation} it follows that 
\begin{equation}
L_{\underline{1}} = e^{2ic} f~, \label{eq:L1}
\end{equation}
where we define $2c$ as the phase of $L_{\underline{1}}$. 

Recall that the Killing spinor equation implies the differential identity \eqref{eq:killing1}, which since $f\neq0$ we rewrite as
\begin{equation}
G = - \frac{1}{4f} \grad K -  \frac{D}{f} \ast K~, \label{eq:killingG}
\end{equation}
which also implies the differential identity \eqref{eq:killing2b} for $\grad f$. We interpret \eqref{eq:killingG} as a relation that determines $G$ in terms of the metric coefficients and $D$. 
In fact, we may expand it explicitly as
\begin{equation}
G = \frac{1}{2} \partial_i f \,  e^{-\sigma} \theta^{\underline{i}} \wedge \theta^{\underline{0}} + \left(\frac{f \rho}{4} - D \right)  \theta^{\underline{1}} \wedge \theta^{\underline{2}}~. \label{eq:killingGb}
\end{equation}
It remains to extract information from \eqref{eq:killing4}.

From the orthonormal frame \eqref{eq:orthoframe} we derive the non-zero spin coefficients as
\begin{align}
\omega_{\underline{0}\underline{0} \underline{i}} &= - e^{-\sigma}\partial_i \ln f \\
\omega_{\underline{i}\underline{0} \underline{j}} &=\frac{1}{2} f \rho\, \epsilon_{ij} \\
\omega_{\underline{0}\underline{i} \underline{j}} &=\frac{1}{2} f \rho\, \epsilon_{ij} \\
\omega_{\underline{i} \underline{j}\underline{k}}  &= 2 e^{-\sigma} \delta_{i[j}\partial_{k]}  \sigma~.
\end{align}
Here we have defined the antisymmetric $\epsilon_{12}\equiv1$ and derivatives on the right-hand side are with respect to the coordinates $x^i$, $i=1,2$. We may then use the spin connection in order to calculate the derivative of $\nabla L$ with coefficients of $L$ given by \eqref{eq:L0}, \eqref{eq:L12relation} and \eqref{eq:L1}, and equate it with the differential Killing bispinor identity in \eqref{eq:killing4}. After some calculation with metric $g$ and field strength $G$ given by respectively \eqref{eq:KKform} and \eqref{eq:killingGb}, the validity of \eqref{eq:killing4} is seen to be equivalent to
\begin{equation}
V+\grad c = - \frac{f^2 \rho}{2} \left( \grad t + A \right)
- \frac{1}{2} \epsilon_{ij} \partial_i \left( \ln f - \sigma\right) \grad x^j~. \label{eq:nablaL}
\end{equation}
We interpret this as a relation that determines $V$ in terms of the metric and the derivative of the phase of $L_{\underline{1}}$. 

A bit more can be said about \eqref{eq:nablaL}. Recall that we have shown that $g$, $G$ and $D$ are all left invariant by $K$, see \eqref{eq:killing0}, \eqref{eq:lieGzero} and \eqref{eq:lieDzero}. The same is true for $V$ up to a gauge transformation. Indeed, from \eqref{eq:nablaL} it follows directly that
\begin{equation}
\lie_K \left( V + \grad c \right) = 0~. \label{eq:killingV}
\end{equation}
In fact, under a $U(1)_R$ gauge transformation \eqref{eq:gaugetrans}, the one-form $L_\mu$ transforms as
\begin{equation}
L_\mu \mapsto e^{-2i\phi} L_\mu~
\end{equation}
and the phase of $L_{\underline{1}}$ transforms as
\begin{equation}
c \mapsto c - \phi~.
\end{equation}
Clearly, $V+\grad c$ is gauge invariant as are the relations in \eqref{eq:nablaL} and \eqref{eq:killingV}. 

We may choose the gauge $c=0$ for which $L_{\underline{1}}=f$ and $L_{\underline{2}}=if$. In this gauge, not only is $L_{\underline{1}}=f$ real and  $L_{\underline{2}}=if$ imaginary, they are both furthermore time-independent. In fact, there is a weaker gauge-fixing condition that may be used,
\begin{equation}
\partial_t c = 0 \Longleftrightarrow V_t = - \frac{f^2\rho}{2}~.
\label{eq:temporal}
\end{equation}
In this gauge the $U(1)_R$ covariant spinor $\epsilon$ is time-independent as well, see \eqref{eq:kstimederivative} in appendix \ref{app:ks}. In any case, in the gauge $c=0$ the orthonormal frame is aligned with the Killing spinor bilinears as $K=f \theta_{\underline{0}}$, $\Re(L)=f\theta^{\underline{1}}$ and $\Im(L)=f\theta^{\underline{2}}$. This however does not imply that $c=0$ is the most appropriate gauge to choose always in order to simplify the solution for $V$.

We summarize that the existence of a timelike Killing spinor implies that the metric takes the form \eqref{eq:KKform}, and both the field strengths $G$ and $\grad V$ are fixed in terms of the metric, as in \eqref{eq:killingGb} and \eqref{eq:nablaL}. The background is locally described by four real scalars $f$, $\rho$, $\sigma$ and $D$ subject to the equations of motion. Similarly to the null case, one need not impose the Killing spinor as a supplementary condition because the equation is integrable, the validity of which statement can be inspected from the equations in appendix \ref{app:ks}. That is, a background subject to  \eqref{eq:KKform},  \eqref{eq:killingGb} and \eqref{eq:nablaL} admits at least locally a timelike Killing spinor and the five scalars are not further constrained by supersymmetry.

We will not write the Killing spinor equation \eqref{eq:KS} explicitly here, for that see the appendix \ref{app:ks}. Rather, we will derive its solution algebraically, which can easily be verified to be satisfying \eqref{eq:KS}. Contracting the Fierz identity \eqref{eq:fierz} with $\epsilon$ from the right and set $\epsilon_1=\epsilon_2=\epsilon$, in order to produce $K \epsilon = i f \epsilon$. In the chosen frame, this becomes
\begin{equation}
\gamma_{\underline{0}} \epsilon = i \epsilon~.
\end{equation}
On the other hand, if we let $\epsilon_2=\epsilon$ and $\epsilon_1=\epsilon^{*}$ and contract with $\epsilon$ from the right then with $\bar{\epsilon}\epsilon^{*}=0$ we are left with $i f \epsilon^{*} = \frac{1}{2} L\epsilon$, which in the chosen frame becomes
\begin{equation}
i \epsilon^{*} =  \frac{1}{2} e^{2 i c} \left( \gamma^{\underline{1}} + i \gamma^{\underline{2}}\right) \epsilon~.
\end{equation}
The rank of the (2,0) supersymmetry spinors is real four-dimensional and the two projection equations determine $\epsilon$ completely as:
\begin{equation}
\epsilon = \sqrt{f} e^{-ic} \epsilon_0~. \label{eq:kssol}
\end{equation}
The spinor $\epsilon_0$ is constant and real in the coordinate system and frame that we use, $\partial_{\mu}\epsilon_0=0$, and satisfies the two projection conditions $\gamma_{\underline{0}}\epsilon=i\epsilon$ and $\gamma_{\underline{1}}\epsilon=i \epsilon^{*}$.

\section{The Model and its Supersymmetric Solutions}
\label{sec:minimal}
We will consider a $N=(2,0)$ model that contains the cosmological constant, the Einstein-Hilbert term, and the gravitational topological term. We will refer to it as $N=(2,0)$ TMG. The lagrangian is given by
\begin{equation}
\begin{aligned}
L &= M \left( 2 D - \epsilon^{abc}C_a G_{bc} \right) \\
&+ R + 4 G_{ab}G^{ab} - 8 D^2 -8 \epsilon^{abc}C_a \partial_b V_c \\
& - \frac{1}{4\mu} \epsilon^{\mu\nu\rho}\left( R_{\mu\nu}{}^{ab} \omega_{\rho ab} +\frac{2}{3} \omega_\mu{}^{ab}\omega_{\nu b}{}^c\omega_{\rho ca} - 8 V_\mu \partial_\nu V_\rho \right)~,\label{eq:lagrangian}
\end{aligned}
\end{equation}
and contains all  $N=(2,0)$ supersymmetry invariant terms up to third-order derivatives \cite{alkac_massive_2015}.  It depends on two real parameters, $\mu$ and $M$. The field equations are
\begin{align}
M&=8D \label{eq:MD}\\
- \grad \ast G &= \frac{1}{2}F + \frac{M}{4} G \label{eq:vector1}\\
F &= 2 \mu \, G \label{eq:vector2}
\end{align}
and the Einstein equation, where $F=\grad V$ and $G=\grad C$. In order to write the Einstein equation, let us define the Cotton tensor
\begin{equation}
C_{\mu\nu} \equiv - \epsilon_\mu{}^{ab}\left( \nabla_bR_{\nu a}-\frac{1}{4}g_{\nu a} \partial_b R\right)~,
\end{equation}
which can be shown to be symmetric and traceless by virtue of the Bianchi identities.
The Einstein equation is
\begin{equation}
R_{\mu\nu} - \frac{1}{2} \left(R+ 2 MD - 8 D^2 \right) g_{\mu\nu} + \frac{1}{\mu} C_{\mu\nu} + 8 G_{\mu a} G_{\nu}{}^a  - 2 G_{ab}G^{ab} g_{\mu\nu}=0~. \label{eq:Einstein}
\end{equation}
Note that neither the vector $V$ nor its field strength $F$ enter the Einstein equation, although \eqref{eq:vector2} makes $F$ to be proportional to $G$. 
The scalar $D$ is likewise fixed algebraically and is in fact rendered constant by \eqref{eq:MD}. 

It should also be noted that a sign of the orientation $\dvol_{g}$ enters the equations of motion in the definition of the Cotton tensor 
 and the Hodge star operator on the left-hand side of \eqref{eq:vector1}. A solution of the equations of motion thus depends on the sign of orientation. 
Furthermore, field equations show that, a solution $(g,V,G,D,\dvol_{g})$ for the theory with parameters $(\mu,M)$ is in one-to-one correspondence with the solution $(g,V,-G,-D,-\dvol_{g})$ for the theory with parameters $(-\mu,-M)$. The Killing spinor equation \eqref{eq:KS} is equivariant under this map by sending $\gamma_{\mu}\mapsto S\gamma^{\mu}S^{-1}=-\gamma_{\mu}$ and the dual solution is still supersymmetric with $\epsilon\mapsto S \epsilon$. 

If we assume a null Killing vector, then $F=0$ implies $G=0$ and the Einstein equation becomes that of $N=1$ cosmological Einstein supergravity with a gravitational topological term. The Killing spinor equation, in which recall $\epsilon$ can be made real, also reduces to that of $N=1$ supergravity.  All null supersymmetric solutions of the theory are thus the same as those found in \cite{gibbons_general_2008}, where all $N=1$ supersymmetric solutions of 
the on-shell topologically massive supergravity are found. They are generically the AdS pp-waves, unless $D=0$ which give the flat space pp-waves. We recall the solutions from \cite{gibbons_general_2008} and also consider here the scaling of the parameters $(M,\mu)$ under a constant scaling of the metric. The supersymmetric waves for $D>0$ are as in \eqref{eq:adsppwave} with unique solution for $h$ up to diffeomorphism invariance
\begin{equation}\label{eq:hnull}
h(u,\rho) = \begin{cases}
e^{(1-\mu\ell)\rho} g(u) &\text{if } \mu\ell\neq 1\\
\rho \, g(u) & \text{if }\mu\ell=1
\end{cases}~,
\end{equation}
where $\ell=(2D)^{-1}=4/M$ is the anti-de Sitter radius. For $D<0$, the supersymmetric solutions are again pp-waves provided one flips the sign of orientation and the sign of $\mu$ in the above solution for the wave profile $h(u,\rho)$. The flat space limit $\ell\rightarrow\infty$ of the supersymmetric pp-wave solution is as in \eqref{eq:gnullDzero} with $h=e^{\mu\rho}g(u)$ for all $\mu\in\RR$ and $M=D=0$.

We will henceforth assume a timelike Killing spinor. The metric is as in \eqref{eq:KKform} and $F$ and $G$ are fixed in terms of the metric:
\begin{align}
F  &= - \frac{1}{2f} \grad \left( f^2\rho\right) \wedge \theta^{\underline{0}} - \frac{1}{2} \left( f^2 \rho^2 + e^{-2\sigma} \partial^2\left( \ln f - \sigma \right) \right) \theta^{\underline{1}}\wedge \theta^{\underline{2}} 
\label{eq:susyF}\\
G &= \frac{1}{2f} \grad f \wedge \theta^{\underline{0}} + \left( 
\frac{\rho f}{4} - D \right) \theta^{\underline{1}}\wedge \theta^{\underline{2}} ~.
\label{eq:susyG}
\end{align}
The last expression implies
\begin{equation}
\grad \ast G = 
\frac{1}{f} \grad \left( \frac{\rho f^2}{4} - Df \right)
\wedge \theta^{\underline{0}} + \left(
\frac{\rho^2 f^2}{4} - D f \rho + \frac{1}{2}e^{-2\sigma}\partial^2 \ln f \right)  \theta^{\underline{1}}\wedge \theta^{\underline{2}} ~,
\end{equation}
where $\partial^2\equiv\sum_{i=1}^2\partial_i\partial_i$. The two vector equations \eqref{eq:vector1} and \eqref{eq:vector2} thus give
\begin{subequations}\label{eq:minimalequs}
\begin{align}
\left(\rho f + 2 \mu\right)f &= \text{const.} \label{eq:const1}\\
e^{-2\sigma}\partial^2 \ln f & =-2 \left( \mu + 2D + \rho f \right) \left( \frac{\rho f}{4} - D \right)~,\label{eq:difff}
\\
-e^{-2\sigma}\partial^2  \sigma &= 2 \left( -\mu + 2D + \rho f \right) \left( \frac{\rho f}{4} - D \right) -f^2\rho^2~. \label{eq:diffsigma}
\end{align}
\end{subequations}
It turns out that the above three equations are sufficient to satisfy the Einstein equation \eqref{eq:Einstein}. We have confirmed this statement 
using a computer algebra system. All timelike supersymmetric solutions of the lagrangian \eqref{eq:lagrangian} are thus given by solutions to \eqref{eq:minimalequs}, 
to which we turn our whole attention. 

\begin{figure}
\begin{center}
\begin{tikzpicture}[
    grow=right,
    level 1/.style={sibling distance=6cm,level distance=5cm},
    level 2/.style={sibling distance=3cm, level distance=6cm},
    edge from parent/.style={very thick,draw=blue!40!black!60,shorten >=5pt, shorten <=5pt},
    edge from parent path={(\tikzparentnode.east) -- (\tikzchildnode.west)},
    kant/.style={text width=2cm, text centered, sloped},
    every node/.style={text ragged, inner sep=2mm},
    punkt/.style={rectangle, rounded corners, shade, top color=white,
   bottom color=blue!50!black!20, draw=blue!40!black!60, very thick }
    ]

\node[punkt] {$f\rho=$const.}
    child {
        node[punkt] {Special Ansatz}
           child { node  [punkt, rectangle split, rectangle split, rectangle split parts=3]          { timelike stretched warped AdS  
\nodepart{second} null  warped AdS \nodepart{third} spacelike squashed warped AdS 
                        } 
                   edge from parent node[kant,below] {$\mu=2D$}
                 }
           child { node [punkt, rectangle split, rectangle split parts=2]          {$\Gamma$-metric
             \nodepart{second}null z-warped
                        } 
                   edge from parent node[kant,above] {$\mu\neq2D$}
                 }
        edge from parent
            node[kant, below] {$f\neq$const.}
          }
    child {
        node[punkt] {General Ansatz}
        child {
            node [punkt,rectangle split, rectangle split,
            rectangle split parts=3] {timelike warped AdS
             \nodepart{second} warped flat
             \nodepart{third} lorentzian sphere
            }
            edge from parent
                node[below, kant,  pos=.6] {twisted case}
        } 
        child {
            node [punkt] {anti-de Sitter
            }
            edge from parent
                node[above, kant] {round case}
        }
            edge from parent{
                node[kant, above] {$f=$const.}}
    };
\end{tikzpicture}
\end{center}
\caption[Ans\"atze tree]{A rooted tree showing the supersymmetric solutions that stem from the constancy Ansatz.}
\label{fig:tree}
\end{figure}
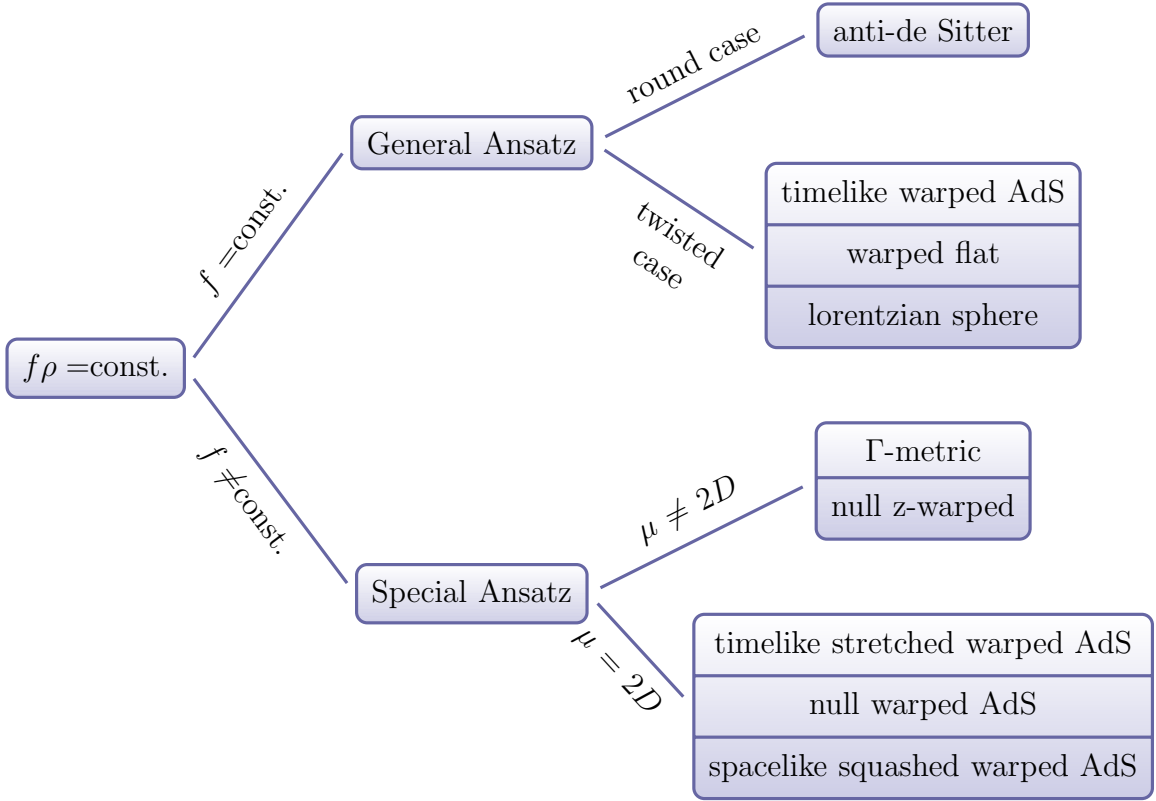

The last equation \eqref{eq:diffsigma} states that the gaussian curvature, which is half of the riemannian scalar curvature, of the two-dimensional metric
\begin{equation}\label{eq:2dmetric}
g_{2d} = e^{2\sigma} \left(  \grad x^2 + \grad y^2 \right)
\end{equation} 
is equal to the right-hand side of \eqref{eq:diffsigma}, that is a given function of $\rho f$, whereas the left-hand side of \eqref{eq:difff} is the respective 2d laplacian acting on $\ln f$ and is equal again to another given function of $\rho f$. By using \eqref{eq:const1} to substitute for $\rho$ in terms of $f$, the two remaining equations \eqref{eq:difff} and \eqref{eq:diffsigma} become two non-linear second-order partial differential equations on $(f,\sigma)$. In what follows, we will solve  \eqref{eq:minimalequs} for
 $(f,\rho,\sigma)$  upon using the \textit{constancy} Ansatz 
\begin{equation}
f\rho = \text{const.} \label{eq:const2}
\end{equation}

In order to motivate this choice, we simply note that if the right-hand side of \eqref{eq:diffsigma} is a constant $\kappa$, then we can 
solve for $\sigma$ uniquely up to 2d diffeomorphisms: depending on the sign of $\kappa$ the two-dimensional metric $g_{2d}$ is either a two-sphere, 
euclidean, or 2d hyperbolic space. We may then easily continue to solve for $f$ from \eqref{eq:difff}. In fact, unless $f \rho = - 2\mu$, then the 
Ansatz \eqref{eq:const2} combined with \eqref{eq:const1} necessitates that $f$ and $\rho$ are separately constant. In that case, the right-hand side 
of \eqref{eq:difff} fixes the constant $\rho f$. 
If on the other hand $f \rho=-2 \mu$, a class of solutions can still be found with $f$ and $\rho$ not necessarily constant. We thus split the 
Ansatz \eqref{eq:const2} into two cases according to whether $f$ is constant or not, which we respectively call the general and the special Ansatz. 
We derive this way many solutions for the metric. Once the metric is known the gauge fields can always be found from \eqref{eq:susyF} and \eqref{eq:susyG}. 
Figure \ref{fig:tree} summarizes all the solutions that we find stemming from the constancy Ansatz \eqref{eq:const2}.

\subsection{General Ansatz}
\label{sec:generic-ansatz}
We take here $f$ and $\rho$ constant so that \eqref{eq:const1} is satisfied and with \eqref{eq:difff} implying either
\begin{subequations}\label{eq:frho}\begin{align}
\rho f &= 4 D  \label{eq:caseround}\\\intertext{or}
\rho f &= -\mu - 2D~. \label{eq:casewarped}
\end{align}\end{subequations}
For reasons that will become apparent from the solutions, we will call \eqref{eq:caseround} the \textit{round case} and \eqref{eq:casewarped} the \textit{twisted case}. 

We still need to solve \eqref{eq:diffsigma}, which we interpret as the statement that the gaussian curvature of $g_{2d}$ is constant and equal to
\begin{equation}
\kappa \equiv 2  \left( -\mu + 2D + \rho f \right) \left( \frac{\rho f}{4} - D \right)- f^2\rho^2~.
\end{equation}
If $\rho f = 4D$ (round case) then 
\begin{subequations}\label{eq:kappa}
\begin{equation}
\kappa = - 16 D^2 ~.
\end{equation}
If on the other hand $\rho f = - \mu-2D$ (twisted case) then
\begin{equation}
\kappa =  2D\left(-2D+\mu\right)~.
\end{equation}
\end{subequations}
In general, $\sigma$ is determined in stereographic coordinates as
\begin{equation}
e^{\sigma} = \frac{1}{1+\frac{\kappa}{4}\left(x^2+y^2\right)}~.
\end{equation}
The two-dimensional metric \eqref{eq:2dmetric} is either a sphere ($\kappa>0$), euclidean space ($\kappa=0$) or hyperbolic space ($\kappa<0$). We can scale time so that $f$ and $\rho$ are scaled but leaving their product invariant. Since $f\neq0$ we choose to scale $f=1$ and then the constant $\rho$ is fixed by \eqref{eq:frho}. It remains to describe the solution in some more familiar setting according to the constant values of $\rho$ and $\kappa$.

\subsubsection{$\kappa<0$}
\label{sec:generalwads}
In this case, the two-dimensional metric is hyperbolic space $H_2$. It is easier to use the conformal coordinates  in which
\begin{equation}
\sigma = - \ln x - \frac{1}{2} \ln |\kappa|~
\end{equation}
and the two-dimensional metric is
\begin{equation}
g_{2d}=\frac{\grad x^2+\grad y^2}{|\kappa|\,x^2 }~.
\end{equation}
Next we integrate  
\begin{equation}
\grad A = \rho\, \dvol_{2d} = \rho\, e^{2\sigma} \grad x^1 \wedge \grad x^2 \label{eq:preintA}
\end{equation}
for a constant $\rho=\rho_0$ and $\sigma$ as above and
\begin{equation}
A = - \frac{\rho_0}{|\kappa|} \frac{1}{x} \grad y~.
\end{equation}
The metric is a Hopf-like fibration over $H_2$
\begin{equation}
g= - \left( \grad t - \frac{\rho_0}{|\kappa|} \frac{1}{x} \grad y \right)^2 + \frac{1}{|\kappa| x^2} \left( \grad x^2 + \grad y^2 \right)~. \label{eq:wadst}
\end{equation}
For $\rho_0^2\neq |\kappa|$ and $\rho_0\neq0$, the spacetime is known as warped timelike anti-de Sitter, which is a homogeneous deformation of anti-de Sitter. 

The isometries of timelike warped anti-de Sitter are generically $
\mathrm{SL}(2,\RR)\times\RR~
$,   
where the $\RR$ is generated by the translation $\partial_t$. The $\mathrm{SL}(2,\RR)$ are the isometries of the $H_2$ base space, which also preserve the total metric. This works as follows.  An infinitesimal isometry $\xi\in\mathfrak{sl}_2$
 of $H_2$, that is $\lie_{\xi} g_{2d} =0$, will preserve the base space volume form, which according to \eqref{eq:preintA} is proportional to $\grad A$. It will thus preserve $A$ up to a suitable compensating time translation 
\begin{equation}
t\mapsto t+ \chi_{\xi}(x,y)~.\label{eq:comptime}
\end{equation}
In so far as the field strength of the (timelike) Kaluza-Klein vector is proportional to the volume-form of the base space, all the volume-form preserving isometries of the base space are lifted to isometries of the total three-dimensional space. There remains a finite isometry that flips the orientation of the two-dimensional base, but that can be compensated in the fiber by a change of coordinates $t\mapsto -t$. An equivalent description of \eqref{eq:wadst} will soon be given in terms of the left-invariant Maurer-Cartan one-forms of $\mathrm{SL}(2,\RR)$.

If $\rho_0=0$ then $A=0$ and \eqref{eq:wadst} is a trivial product $\RR_t \times H_2$,
\begin{equation}\label{eq:wadstdegenerates}
g = - \grad t^2 +  \frac{1}{|\kappa| x^2} \left( \grad x^2 + \grad y^2 \right)~.
\end{equation} The isometries are still $\mathrm{SL}(2,\RR)\times\RR$ but without the compensating time translation \eqref{eq:comptime}. This happens only in the twisted case \eqref{eq:casewarped} and in particular when $\mu=-2D$ so that $\kappa=-8D^2$. 

If $\rho_0^2 = |\kappa|$ then there is an enhancement of symmetry and in fact \eqref{eq:wadst} becomes the metric of three-dimensional anti-de Sitter of radius $2/\sqrt{|\kappa|}$ and with isometry $
\mathrm{SL}(2,\RR)_L \times \mathrm{SL}(2,\RR)_R$~. To see how the enhancement of isometry comes about, we may write the left-invariant Maurer-Cartan one-forms of $\mathrm{SL}(2,\RR)$ as
\begin{subequations}\label{eq:slframes}
\begin{align}
\tau_0 &= \grad t - \frac{1}{x} \grad y\\
\tau_1 &=  \frac{1}{x} \cos t \, \grad x+ \frac{1}{x}\sin t \, \grad y \\
\tau_2 &=  -\frac{1}{x} \sin t \, \grad x + \frac{1}{x}\cos t \, \grad y  ~.
\end{align}\end{subequations}
Then timelike warped anti-de Sitter \eqref{eq:wadst} is 
\begin{equation}
 g= \frac{1}{|\kappa|} \left( - \frac{\rho_0^2}{|\kappa|} \tau_0^2 +  \tau_1^2 +  \tau_2^2 \right) \label{eq:wadst1}
\end{equation}
and is manifestly invariant under the left action $\mathrm{SL}(2,R)_L$ and a translation $\partial_t \in\mathrm{SL}(2,R)_R $ that preserves separately $\tau_0^2$ and $\tau_1^2 +  \tau_2^2$, see for instance how $\partial_t$ acts on the  Maurer-Cartan one-forms in \eqref{eq:slframes}. When  $\rho_0^2 = |\kappa|$, then the bilinear in 
 \eqref{eq:wadst2} is the Killing form on $\mathfrak{sl}(2,\RR)$
\begin{equation}
 g= \frac{1}{|\kappa|} \left( - \tau_0^2 +  \tau_1^2 +  \tau_2^2 \right) \label{eq:roundads}
\end{equation}
 and the metric is invariant under both the left and the (now full) right action. For various coordinate descriptions of three-dimensional anti-de Sitter see the appendix D of \cite{mitsuka_no_2012} and for the coordinates of \eqref{eq:slframes} see in particular equation (126) there. The bi-invariant metric on $\mathrm{SL}(2,\RR)$ is that of three-dimensional anti-de Sitter.

Let us find when the ``round'' anti-de Sitter is possible, that is to find when $\rho_0^2=|\kappa|$ holds. This is clearly always the case for the round case \eqref{eq:caseround}, which also always gives $\kappa<0$ provided $D\neq0$. The twisted case \eqref{eq:casewarped} is more restrictive. Round anti-de Sitter is possible in the twisted case either when $\mu+6D=0$, but then the twisted case \eqref{eq:casewarped} coincides with the round case  \eqref{eq:caseround}, or when $\mu=0$ so that $|\kappa|=4D^2$. However the $\mu=0$ possibility is also degenerate in a sense, since the contribution of the Cotton tensor to the Einstein equation is weighted by a factor of ${\mu}^{-1}$. We summarize by saying that we have so far found two classes of solution: either anti-de Sitter provided $D\neq0$, or timelike warped anti-de Sitter provided 
\begin{equation}\label{eq:negkappa}
\kappa = 2D \left( -2 D + \mu \right) < 0~.
\end{equation}
The two parameters of  timelike warped anti-de Sitter are parametrized by $(\mu,D)$. When $2D+\mu=0$, the timelike warped anti-de Sitter family degenerates into a direct product $\RR_t \times H_2$ and when $\mu+6D=0$, the family degenerates into anti-de Sitter.

\subsubsection{$\kappa=0$}
\label{sec:kappazero}
In this case the two-dimensional base is euclidean. Inspecting \eqref{eq:kappa}, the case $\kappa=0$ happens if either
\begin{enumerate}[a)]
\item $D=0$ and $\rho_0=0$ (round case \eqref{eq:caseround}),
\item  $D=0$ and $\rho_{0}=-\mu$ (twisted case \eqref{eq:casewarped}), or 
\item $\mu=2D$ and $\rho_0=-4D$ (twisted case \eqref{eq:casewarped}). 
\end{enumerate}
The metric is generically given by
\begin{equation}
g = - \left( \grad t + \rho_0 \, x \,\grad y \right)^2 + \grad x^2 + \grad y^2~. \label{eq:eucwarp}
\end{equation}
Case a) manifestly gives Minkowski space. It can be seen as the limit of the round case \eqref{eq:caseround} with $D\neq0$. The other two cases can also be seen as the $\kappa\rightarrow0$ limits of the timelike warped anti-de Sitter \eqref{eq:negkappa}. Finally, we note that case c) at $D=0$ degenerates to case a).

In the twisted case, $\rho_0\neq 0$, the isometries are generically $
\mathrm{Iso}(\RR^2) \times \RR$ 
where the isometries of the two-dimensional euclidean metric are
$
\mathrm{Iso}(\RR^2) = \mathrm{O}(2) \ltimes \RR^2~
$ 
and can be always lifted to preserve the fiber, that is the Kaluza-Klein connection, with a suitable compensating shift of $t$. 
The metric \eqref{eq:eucwarp} has appeared before in \cite{moutsopoulos_homogeneous_2013} as the solution called there EC11.
We may call this metric in analogy to timelike warped anti-de Sitter as timelike warped flat. If $\rho_{0}=0$ the isometries enhance to the Poincar\'e group of three-dimensional Minkowski space.

\subsubsection{$\kappa>0$}
By inspecting \eqref{eq:kappa}, $\kappa>0$ can only happen in the twisted case. We rewrite the conditions again
\begin{align}
\rho_0 &= -\mu - 2D~,\\
\kappa &=  2D\left(-2D+\mu\right) > 0~.
\end{align}
It follows that $\rho_0$ can never be zero. The metric may be written as a fibration over a sphere
\begin{equation}
g = - \left( \grad t + A \right)^2 + \frac{1}{\kappa} \left( \grad \theta^2 + \sin^2\theta \grad \phi^2 \right)
\label{eq:lhopf}
\end{equation}
where we integrate $A$ from
\begin{equation}\label{eq:dAlor}
\grad A = \rho \, \dvol_{2d} = \frac{\rho_0}{\kappa} \sin\theta \, \grad \theta \wedge  \grad \phi
\end{equation}
into
\begin{equation}\label{eq:Alor}
A= - \frac{\rho_0}{\kappa}  \cos \theta \, \grad \phi~.
\end{equation}
The metric \eqref{eq:lhopf} is a lorentzian signature version of the Hopf fibration of the sphere. It is more familiar as the $SU(2)$-orbit, the geometry at constant Schwarzschild radius, of the four-dimensional Taub-NUT space \cite{misner_taub-nut_1965,moutsopoulos_nut_2010}. We recall from \cite{misner_taub-nut_1965} that the lorentzian sphere \eqref{eq:lhopf} and \eqref{eq:dAlor} necessarily has the topology of $S^3$ and thus also closed timelike curves. 

The isometries of \eqref{eq:lhopf} are manifestly $
\mathrm{O}(3) \times \RR$. 
As before, we may argue for this because $\grad A$ is the volume form of the base space two-sphere (up to some constant) and thus $A$ is invariant under a base space infinitesimal isometry $\xi\in\mathfrak{su}(2)$, provided that we compensate with a suitable time shift
\begin{equation}
t \mapsto t+ \chi_{\xi}(\theta,\phi)~.
\end{equation} 
The parity $\ZZ_2\subset \mathrm{O}(3)$ is provided by an isometry of the two-sphere that does not preserve the two-dimensional orientation, compensated by a parity transformation on time $t\mapsto -t$. We may write the left-invariant Maurer-Cartan one-forms of $\mathrm{SU}(2)$ in the coordinates of the Hopf fibration
\begin{align}
\sigma_1 &=  \grad \theta \cos t  + \sin \theta \sin t \grad \phi \\
\sigma_2 &=  -\grad \theta \sin t  + \sin \theta \cos t \grad \phi \\
\sigma_3 &= \grad t - \sin\theta \grad \theta \grad \phi
\end{align}
so that the metric becomes an $\mathrm{SU}(2)$ invariant metric on the sphere, seen as the group manifold $\mathrm{SU}(2)$,
\begin{equation}
g = -  \frac{\rho_0^2}{\kappa^2} \sigma_3^2 + \frac{1}{\kappa}\left( \sigma_1^2 + \sigma_2^2 \right)~. \label{eq:defsphere}
\end{equation}
More generally, left-invariant metrics on $\mathrm{SU}(2)$ can be diagonalized in the left-invariant basis by use of the adjoint action on $\mathrm{SU}(2)$, so one is left with three parameters describing the metric. This is the usual Bianchi IX form of an  $\mathrm{SU}(2)$-invariant metric. In our case two of the diagonal parameters are the same and positive but the third is negative. That is, the metric may be called a biaxial deformation\footnote{Strictly speaking, one would not be able to deform continuously from the sphere into our metric while preserving non-degeneracy because the metric changes signature and its determinant changes sign.} of the round three-sphere, albeit in an unusual lorentzian signature. 

Let us conclude section \S\ref{sec:generic-ansatz} with an observation regarding the gauge field $V$. Recall that the gauge invariant strength is fixed by supersymmetry, see \eqref{eq:susyG} from where it follows that in the twisted case it is non-zero and, regardless of the sign of $\kappa$, equal to
\begin{equation}
G = - \frac{6 D + \mu}{4} \theta^{\underline{1}}\wedge \theta^{\underline{2}}~.
\end{equation}
For instance, one reads that the vector $\ast G$ is always timelike and of constant norm. The vector $V$ is also fixed by supersymmetry as in \eqref{eq:nablaL}, the right-hand side of which is in fact gauge invariant. However, choosing $c=0$ is not the most appropriate choice to solve for $V$. Rather, one can find another gauge choice for $c$ such that $V$ is always timelike and with constant norm. 
Furthermore, in this gauge $V$ is left invariant by all isometries of the metric. 
\subsection{Special Ansatz}
\label{sec:special-ansatz}
We now consider the case
\begin{equation}
f \rho = -2 \mu~,
\end{equation}
but require a solution where $f$ is not constant. We thus consider non-trivial solutions to
\begin{align}
e^{-2\sigma} \partial^2 \ln f &= {\kappa}_2 \equiv 4D^2-\mu^2\\
e^{-2\sigma} \partial^2 \sigma &= \kappa_1 \equiv (2D-\mu)^2~.
\end{align}
If $\kappa_1\neq0$, then the second equation is solved uniquely up to two-dimensional diffeomorphisms by
\begin{equation}
\sigma = - \frac{1}{2}\ln\kappa_1 - \ln x~.
\end{equation}
The first equation then has general solution
\begin{equation}
f = e^{h} x^{-\frac{{\kappa}_2}{\kappa_1}}~,
\end{equation}
where $h$ is a harmonic function on $H_2$. For simplicity we will let $h$ depend only on $x$ and thus
\begin{equation}
h = { c_1 }\, x + d_1~.
\end{equation}
By a scaling of time, we may set $d_{1}=0$. Whatever choice we make for $h$, the solution of $\rho$ is
\begin{equation}
\rho = - 2\mu\, e^{-h} x^{\frac{{\kappa}_2}{\kappa_1}}~.
\end{equation}
If $\kappa_1=0$, which only happens when $\mu=2D$, the general solution is $\sigma=0$, $f=e^h$ and $\rho=-2\mu e^{-h}$, where $h$ is harmonic on $\RR^2$. Again, we will choose for simplicity $h={ c_1 }\,x$. It remains to unravel the geometries of these solutions.

\subsubsection{$\kappa_1\neq0$}
We first investigate $\kappa_1\neq 0$ and ${ c_1 }\neq0$. We will only consider the case ${ c_1 }>0$. The solution to
\begin{equation}
\grad A = \rho \, e^{2\sigma} \grad x \wedge \grad y = - \frac{2\mu}{\kappa_1} e^{-{ c_1 }x} x^{\frac{\kappa_2}{\kappa_1}-2}  \grad x \wedge \grad y \label{eq:gradA1}
\end{equation}
is
\begin{equation}
A= \frac{2\mu}{\kappa_1}  { c_1 }^{1-\frac{\kappa_2}{\kappa_1}} \Gamma\left(\frac{\kappa_2}{\kappa_1}-1, { c_1 } x \right) \grad y~,
\end{equation}
where
\begin{equation}
\Gamma(s,x) = \int x^{s-1} e^{-x} \grad x
\end{equation}
is a generalized gamma function\footnote{The singularities of $\Gamma(s,x)$ are at $x=0$ and at non-positive integers $s$.}. In fact we may absorb the factor of ${ c_1 }$ in $x$, $y$ and $t$ and write the metric as
\begin{equation}\label{eq:gammametric}
g = - e^{2x} x^{-2 \frac{\kappa_2}{\kappa_1}} \left( \grad t + 
\frac{2\mu}{\kappa_1}  \Gamma\left(\frac{\kappa_2}{\kappa_1}-1, x \right) \grad y
\right)^2 + \frac{\grad x^2 + \grad y^2}{\kappa_1 \, x^2}~.
\end{equation}
Due to the appearance of the gamma function, we shall refer to this solution as the $\Gamma$-metric. We are not aware of any special properties of the  $\Gamma$-metric, or if it has appeared elsewhere before. We present here two 
curvature invariants of the $\Gamma$-metric as a function of $x$:
\begin{align}
R &= -24 D^2 -2 \kappa_1 x^2+4 \kappa_2 x
\\
R_{\mu\nu}R^{\mu\nu} &=
192D^4
+2 \kappa_1^2 x^4
+ \left(2 \kappa_1^2-8 \kappa_1 \kappa_2 \right)x^3 \\&
+ \left(2 \kappa_1^2+2 \kappa_1 \kappa_2+10  \kappa_2^2\right)x^2
-2\kappa_2(3\kappa_1 + 5\kappa_2 + 16D\mu) x \nonumber
\end{align}
The case ${ c_1 }<0$ might likewise be interesting.

Next, we investigate $\kappa_1\neq0$ and ${ c_1 }=0$. Recall, we already have $f=x^{-\frac{{\kappa}_2}{\kappa_1}}$. We also take $\kappa_1\neq\kappa_2$, which is equivalent to $\mu\neq0$. We readily integrate \eqref{eq:gradA1} so that
\begin{equation}
A = - \frac{2\mu}{\kappa_2-\kappa_1} x^{\frac{\kappa_2}{\kappa_1}-1} \grad y
 ~.
\end{equation}
The metric can then be rewritten as
\begin{equation}
g = \frac{1}{\left(2D-\mu\right)^2}\left(- x^{-2 \frac{\kappa_2}{\kappa_1}} \left( \grad t +
 x^{ \frac{\kappa_2}{\kappa_1}-1} \grad y
\right)^2 + \frac{\grad x^2 + \grad y^2}{x^2}\right)~,\label{eq:prepp}
\end{equation}
where we took out a factor of $\kappa_1=\left( (\kappa_2-\kappa_1)/(2\mu)\right)^2 = (2D-\mu)^2$. For $D\neq0$ (thus $\kappa_1\neq-\kappa_2$) we can bring this to a more familiar form.  Later we will also return to the $D=0$ case. For $\mu=-2D$ we have $\kappa_2=0$ and the space is anti-de Sitter, see \eqref{eq:roundads}, so let us assume $\kappa_2\neq0$. If we change to coordinates
\begin{equation}
u = \left( 1+ \frac{\kappa_2}{\kappa_1}\right)^{-1}  t, \quad  v =-  \left( 1+ \frac{\kappa_2}{\kappa_1}\right)^{-1} y, \quad  w= x^{- \frac{\kappa_2}{\kappa_1}-1} 
\end{equation}
the metric becomes
\begin{equation}
g = \frac{16D^2}{\left(2D-\mu\right)^4}\left(- w^{2z}  \grad u^2  + \frac{\grad w^2 + 2 \grad u \grad v}{w^2}\right)~,  \label{eq:zwarped}
\end{equation}
where $z\equiv\kappa_2/(\kappa_1+\kappa_2)$. 
We recognize this metric as a AdS pp-wave, whose general form up to the cosmological scale was given previously in \eqref{eq:adsppwave}.  One can easily calculate that $\ast G$ is everywhere null as we expect for a pp-wave, unlike $\ast G$ of the $\Gamma$-metric solution \eqref{eq:gammametric}. 

The metric \eqref{eq:zwarped} enjoys a non-relativistic scaling symmetry, 
\begin{equation}
u \mapsto \Lambda^{-z} u , \quad w \mapsto \Lambda w, \quad v \mapsto \Lambda^{2+z} v~.
\end{equation}
 This property has led to research in the area of holography, e.g. in \cite{anninos_curious_2010} where it was called null $z$-warped. Interestingly, here, it appears as a ${ c_1 }=0$ limit of the previous $\Gamma$-metric. If furthermore $z=-2$ then the spacetime is a $\mathrm{SL}(2,\RR)\times\RR$ invariant metric on $\mathrm{SL}(2,\RR)$, which can be made lucid in terms of the left-invariant Maurer-Cartan one-forms
\begin{equation}
g =  \frac{64D^2}{\left(2D-\mu\right)^4} \left( - \tau_0^2 + \tau_1^2 + \tau_2^2 - \left( \tau_0+\tau_1 \right)^2 \right)~\label{eq:znull}
\end{equation}
in some coordinate system. Note that by an automorphism of $\mathrm{SL}(2,\RR)$, in particular a lorentzian boost that will scale $\tau_0+\tau_1 $ and leave invariant the rest in \eqref{eq:znull}, the deformation of the metric by $\left( \tau_0+\tau_1 \right)^2$ is sensitive only on the sign of the deformation and not its scale: our case is null warped anti-de Sitter with the minus sign deformation. For $z\neq2$ it generically enjoys only its manifest symmetries: non-relativistic scalings, $u$-shifts and $v$-shifts. 

If $D=0$ and thus $\kappa_1=-\kappa_2$, we still get a pp-wave metric from \eqref{eq:prepp}, but now a flat space pp-wave. Indeed, by defining $w = \ln x$ the metric  \eqref{eq:prepp} becomes after rescaling of $t$, $y$ and $w$ by factors of $\mu$:
\begin{equation}\label{eq:zflat}
g= - e^{2w/\mu} \grad t^2 - 2 \grad t \grad y + \grad w^2~.
\end{equation}
Similar to null warped, the deformation $e^{2w}\grad t^2$ does not depend on its size, but it does depend on its sign. Here again, we find a supersymmetric deformation with a negative sign of deformation.

\subsubsection{$\kappa_1=0$}
\label{sec:specialwads}
It remains to investigate the case $\kappa_1=0$, which is equivalent to $\mu=2D$ and gives $\kappa_2=0$ as well. Assume first $h={ c_1 }\, x$ with ${ c_1 }\neq0$. Similarly to our previous procedure, we integrate $\grad A$ in terms of $\rho=-2\mu e^{-{ c_1 } x}$ and find the metric
\begin{equation}
g = - e^{2 { c_1 } x} \left( \grad t + \frac{2\mu}{{ c_1 }}e^{-{ c_1 }\, x} \grad y\right)^2 + \grad x^2 + \grad y^2 ~.
\end{equation}
At this point we may restrict to the choice ${ c_1 }>0$ and $\mu>0$. If this is not so, we use the parity transformations $x\mapsto -x$ and / or $(t,y)\mapsto(-t,-y)$ to achieve so in the form of the metric. A change of coordinates 
\begin{equation}
w = e^{{ c_1 }x} , \quad t' = { c_1 } t, \quad y' = { c_1 } y
\end{equation} 
(and immediately dropping the primes) will be useful in order to bring an overall ${ c_1 }^{2}$ out of the metric:
\begin{equation}
g= \frac{1}{{ c_1 }^2} \left( \frac{\grad w^2}{w^2} + \left( 1- \frac{4\mu^2}{{ c_1 }^2} \right) \grad y^2 - w^2 \grad t^2 - \frac{4\mu}{{ c_1 }} w \grad t \grad y \right) ~.\label{eq:manyfaces}
\end{equation}
The metric can be made more familiar, but the ensuing discussion depends on the sign of ${ c_1 }-2\mu$, that is whether $g_{yy}$ above is positive or negative. 

If ${ c_1 }>2\mu$ we scale the coordinates with
\begin{align}
u &= \sqrt{\frac{1-4\mu^2/{ c_1 }^2}{4\mu^2/{ c_1 }^2}} y \\
\tau & = \frac{1}{\sqrt{1-4\mu^2/{ c_1 }^2}} t~,
\end{align}
and the metric \eqref{eq:manyfaces} takes the form of
\begin{equation}\label{eq:wadss}
g = \frac{1}{{ c_1 }^2}\left(\frac{\grad w^2}{w^2} - w^2 \grad \tau^2 + \frac{4\mu^2}{{ c_1 }^2} \left(  \grad u + w \grad \tau \right)^2 \right)~.
\end{equation}
We recognize this metric as spacelike warped anti-de Sitter in so-called ``extremal accelerating'' coordinates, with a ``squashed'' deformation $4\mu^2/{ c_1 }^2<1$. Such coordinates were described in \cite{jugeau_accelerating_2011}. In terms of the Maurer-Cartan one-forms on $\mathrm{SL}(2,\RR)$
\begin{subequations}\label{eq:MCspace}
\begin{align}
\tau_0 &= w \cosh u \, \grad \tau +  \frac{\sinh u}{w} \grad \tau\\
\tau_1 &= \frac{\cosh u}{w} \grad \tau +   w \sinh u \, \grad \tau\\
\tau_2 &=  \grad u + w \grad \tau~.
\end{align}\end{subequations}
the metric is
\begin{equation}\label{eq:wadsstaumet}
g =  \frac{1}{{ c_1 }^2}\left( - \tau_0^2 + \tau_1^2 + \frac{4\mu^2}{{ c_1 }^2} \tau_2^2 \right)
\end{equation}
and enjoys $\mathrm{SL}(2,\RR) \times \RR$ symmetry. If ${ c_1 }<2\mu$ we scale the coordinates similarly to as before
\begin{align}
u &= \sqrt{\frac{4\mu^2/{ c_1 }^2-1}{4\mu^2/{ c_1 }^2}} y \\
\tau & = \frac{1}{\sqrt{4\mu^2/{ c_1 }^2-1}} t~,
\end{align}
and the metric  \eqref{eq:manyfaces} takes the form of
\begin{equation}\label{eq:wadst2}
g = \frac{1}{{ c_1 }^2}\left(\frac{\grad w^2}{w^2} + w^2 \grad \tau^2 - \frac{4\mu^2}{{ c_1 }^2} \left(  \grad u + w \grad \tau \right)^2 \right)~.
\end{equation}
We recognize this metric as timelike warped anti-de Sitter, which was described earlier in \eqref{eq:wadst} and \eqref{eq:wadst1}. The deformation from anti-de Sitter is here necessarily ``stretched'', that is $4\mu^2/{ c_1 }^2>1$. Finally, if $2\mu={ c_1 }$, the metric \eqref{eq:manyfaces} becomes after $w\mapsto w^{-2}$ and scaling $t$ appropriately
\begin{equation}\label{eq:wadsn}
g= \frac{4}{{ c_1 }^2}\left( \frac{\grad w^2+\grad t \grad y}{w^2} - w^{-4} \grad t^2 \right)~.
\end{equation}
This is the metric of null warped anti-de Sitter, which was described earlier as a special case of null $z$-warped, see around \eqref{eq:zwarped} (special Ansatz $\kappa_1\neq0$ with ${ c_1 }=0$ and set $z=-2$).

The constant $c_1\neq0$ is arbitrary but characterizes the solutions of this section completely. As we have seen, the sign of $c_1-2\mu$ differentiates between spacelike \eqref{eq:wadss}, timelike \eqref{eq:wadst2} and null \eqref{eq:wadsn} warped anti-de Sitter. The magnitude of $c_1$ also characterizes the \textit{effective} anti-de Sitter radius and the size of the deformation. The constant $c_1$ also  characterizes the norm of $V^2$ and $G^2$,
\begin{equation}
V_{\mu}V^{\mu}=\frac{1}{2}G_{\mu\nu}G^{\mu\nu}= \frac{{c_{1}}^2-4\mu^2}{4} ~,
\end{equation} 
in the gauge $c=0$ for $V$, see \eqref{eq:nablaL}. 

It may appear that we have neglected one case, the case $\kappa_1=0$ with $h=0$. This case yields the metric
\begin{equation}
g = - \left( \grad t -2 \mu\,  x \, \grad y \right)^2 + \grad x^2  + \grad y^2~.
\end{equation}
This is a case of the timelike warped flat metric that we came across earlier in \eqref{eq:eucwarp}, when investigating the general Ansatz for $\kappa=0$. It has both $f$ and $\rho$ constant so it is included in the general Ansatz of section \S\ref{sec:generic-ansatz}. In any case, it is seen here as a ${ c_1 }=0$ limit of the previous $\kappa_1 =0$ solutions, under which the warped anti-de Sitter limits to what we call warped flat.

\section{Supersymmetric Black Holes from Quotients}
\label{sec:asquotients}
The regular isometric quotients of spacelike, timelike and null warped anti-de Sitter were investigated in \cite{anninos_warped_2009}. One requires either a quotient with no pathologies or if such pathologies arise (in particular closed causal curves) that they are hidden behind an absolute horizon \`a la the BTZ construction. The supersymmetric warped anti-de Sitter solutions that we find in the theory are: timelike (both stretched and squashed) in \eqref{eq:wadst}, timelike stretched in \eqref{eq:wadst2}, spacelike squashed in \eqref{eq:wadss}, and null warped with negative deformation in \eqref{eq:zwarped} for $z=-2$ and in \eqref{eq:wadsn}. Scanning the results of \cite{anninos_warped_2009} and comparing
with our solutions, the quotients that we
are interested in, namely those that have
black hole like properties, correspond to
\begin{itemize}
\item The self-dual spacelike anti-de Sitter quotient. That is \eqref{eq:wadss} with $u = T\, \theta$ and $\theta=\theta+2\pi$ a periodic angle,
\begin{equation}
g = \frac{1}{{ c_1 }^2}\left(\frac{\grad w^2}{w^2} - w^2 \grad \tau^2 + \frac{4\mu^2}{{ c_1 }^2} \left( T \grad \theta + w \grad \tau \right)^2 \right)~. \label{eq:selfdual}
\end{equation}
In the notation\footnote{The left-invariant vector fields $r_a$ satisfy the $\mathfrak{sl}_2=\mathfrak{so}(1,2)$ algebra $[r_a,r_b]=\epsilon_{ab}{}^c l_c$ for a ``lorentzian'' symbol $\epsilon_{abc}$, $a=0,1,2$, and similarly for the right-invariant $l_a$. In other nomenclature, the combination $l_0\pm l_2$ is parabolic in $\mathfrak{sl}_2$, $l_0$ is elliptic (compact) and $l_2$ is hyperbolic (non-compact).}
 of \cite{jugeau_accelerating_2011} the closed isometry is generated by $\partial_{\theta} =T\, l_2$.
\item The quotient corresponding to timelike squashed anti-de Sitter in  \eqref{eq:wadst} where we identify $y=\theta$ with $\theta=\theta+2\pi$:
\begin{equation}
g= - \left( \grad t - \frac{\rho_0}{|\kappa|} \frac{1}{x} \grad \theta \right)^2 + \frac{1}{|\kappa| x^2} \left( \grad x^2 + \grad \theta^2 \right)~.\label{eq:halfvac}
\end{equation}
In the notation of \cite{jugeau_accelerating_2011} the closed isometry is generated by $\partial_{\theta} = r_0+r_2$. 
Inserting a period $T$ here is not physical, since it can be absorbed by a rescaling of $x$. The timelike fiber is squashed when $\rho_0^2<|\kappa|$ that, including the trivial fiber $\RR \times H_2$ for $\rho_0=0$, is the condition $D\neq0$ with
\begin{equation}
 \frac{\mu}{D} \in \left( - 6, 0\right)~.
\end{equation}
\end{itemize}
In these two cases, the closed curves generated by $\exp\left(\zeta\partial_{\theta}\right)$, $\zeta\in[0,2\pi]$, are always spacelike and there are no causal pathologies. 

The Killing horizon $w=0$ in the self-dual quotient is apparent and global coordinates exist that remove the coordinate singularity. We may see this without using a coordinate transformation as follows. We first assert that the two-dimensional base metric is $\AdS_2$ and so a two-dimensional coordinate transformation $(w,\tau)\mapsto (\sigma,t)$ exists such that 
\begin{equation}\label{eq:2dmetricsame}
\frac{\grad w^2}{w^2} - w^2 \grad \tau^2 = \grad \sigma^2 - \cosh^2\sigma \, \grad t^2~.
\end{equation}
For this change of coordinates, the two-dimensional volume form, which recall is proportional to $\grad A$, becomes
\begin{equation}\label{eq:2dvolsame}
\grad \left( w \, \grad \tau\right) = \grad w \wedge \grad \tau = \cosh \sigma \, \grad \sigma \wedge \grad t = \grad\left( \sinh\sigma \, \grad t\right)~.
\end{equation}
From \eqref{eq:2dmetricsame} and \eqref{eq:2dvolsame} we deduce that a compensating gauge shift 
\begin{equation}\label{eq:sdiff}
\theta'=\theta+\chi(w,\tau)
\end{equation}should exist that brings the three-dimensional metric \eqref{eq:selfdual} to the form
\begin{equation}\label{eq:selfdual2}
g = \frac{1}{{ c_1 }^2}\left(  \grad \sigma^2 - \cosh^2\sigma \, \grad t^2  + \frac{4\mu^2}{{ c_1 }^2} \left( T \grad \theta' + \sinh\sigma \grad t \right)^2 \right)~.
\end{equation}
An explicit diffeomorphism implementing \eqref{eq:2dmetricsame} and \eqref{eq:sdiff} can be found in \cite{deger_supersymmetric_2013} and is unique up to isometries. In the coordinates of \eqref{eq:selfdual2}, the identification is $\theta'=\theta'+2\pi$. Also note that the diffeomorphism that brings \eqref{eq:selfdual} to \eqref{eq:selfdual2} acts on the boundary of spacetime, which may hence act non-trivially in the quantum theory\footnote{In fact, the coordinates of \eqref{eq:selfdual} only describe part of the conformal boundary, see \cite{spradlin_vacuum_1999} or \cite{jugeau_accelerating_2011}.}, see e.g. \cite{hawking_soft_2016}. Accordingly, the self-dual quotient in the coordinates of \eqref{eq:selfdual} was interpreted in \cite{anninos_warped_2009} as a black hole, see also \cite{spradlin_vacuum_1999}.

It is easy to see that for both metrics \eqref{eq:selfdual} and \eqref{eq:halfvac}, provided that the gauge choice $c$ does not depend on $\theta$, the Killing spinor $\epsilon$ as solved for in \eqref{eq:kssol} also does not depend on $\theta$. Therefore, at least one complex supersymmetry is preserved by the identification $\theta=\theta+2\pi$ in the two quotients. 

Let us finally review the supersymmetric quotients of anti-de Sitter as known from \cite{coussaert_supersymmetry_1994}. They are the (now undeformed) spacelike self-dual solution \cite{coussaert_selfdual_1994}, that is \eqref{eq:selfdual} but with replacing $2\mu={ c_1 }=4D$, the (massless) BTZ vacuum
\begin{equation}\label{eq:btzvac}
g = \ell^2 \frac{\grad r^2 - \grad t^2 + \grad \theta^2}{r^2}~,
\end{equation}
which corresponds in  the notation of \cite{jugeau_accelerating_2011} to $\partial_{\theta}=r_0+r_2+l_2+l_0$, and finally the extremal BTZ black holes
\begin{equation}
g= - N(r)^2 \grad t^2 + \frac{\grad r^2}{N(r)^2} + r^2 \left( \grad \theta + \frac{a^2}{\ell r^2} \grad t\right)^2 
\end{equation}
with 
\begin{equation}
N(r)=\frac{r^2-a^2}{\ell r}~.
\end{equation}
The identification in the latter black hole  in  the notation of \cite{jugeau_accelerating_2011} is $\partial_{\theta} = r_2+r_0+ a \, l_2$. Its mass $m$ and angular momentum $j$ saturate the extremal bound $|j|=m \ell=2 a^2/\ell$ and $r=a$ is an extremal horizon. Let us note that the equivalent warped black hole is not in our solution space, because the equivalent BTZ quotient as studied in \cite{anninos_warped_2009} involves a spacelike stretched warping. Conversely, the timelike squashed quotient in \eqref{eq:halfvac} is not regular for the round anti-de Sitter: it contains closed null curves.

\section{Discussion}
In this paper, after studying implications of the Killing spinor equation of the off-shell (2,0) supergravity, we applied our findings 
to $N=(2,0)$ TMG. The null case reduces immediately to on-shell $N=(1,0)$ TMG whose solutions are pp-waves \cite{gibbons_general_2008}.
In the timelike case, we imposed the {\it constancy} ansatz \eqref{eq:const2} on metric functions which lead to a large number of solutions which are 
summarized in the Table \ref{table}. This ansatz is weaker than the condition used for 
the $N(1,1)$ TMG in \cite{deger_supersymmetric_2013} where it was assumed that the auxiliary vector field is constant in a flat basis which
lead to only warped AdS solutions. It would be interesting to see if the {\it constancy} ansatz \eqref{eq:const2} or some modification of it
would give more solutions in the $N(1,1)$ case. Another challenge is to find additional supersymmetric solutions in  $N=(2,0)$ TMG by relaxing 
the {\it constancy} requirement. 

The conditions for an off-shell (2,0) background to be maximally supersymmetric were analyzed in \cite{knodel_rigid_2015} without referring to a particular model.  
In our case, Minkowski and anti-de Sitter, which both have $G=0$, are maximally supersymmetric as expected. If $G\neq 0$, the conditions of  
\cite{knodel_rigid_2015}  reduce in our model to $\mu=2D$ and the equation
\begin{equation}
R_{\mu\nu} = -8 D^2 g_{\mu\nu} +4 G_{\mu a}G_{\nu}{}^a~.
\end{equation}
This is satisfied for all $\mu=2D$ solutions that we have found, namely the warped anti-de Sitter special Ansatz solutions of section \S\ref{sec:specialwads} and 
the warped flat case (c) of the general Ansatz in section \S\ref{sec:kappazero}. From the Killing spinor equation \eqref{eq:KS} it is easy 
to see that if $\epsilon$ is a Killing spinor then so is $i\epsilon$.  Hence, we conclude that our remaining solutions should be half supersymmetric. 

The solution that surprised us the most was \eqref{eq:gammametric} where the Gamma function shows up in the metric. As far as we know, such a solution have not 
appeared before and its properties needs further investigation. 

Another future direction is to consider a more general off-shell $N=(2,0)$ model that includes higher order invariants that were constructed 
in \cite{alkac_massive_2015}. In the  analogous situation for the $N=(1,1)$ case, warped AdS solutions of the minimal model survived with shifted 
parameters and some extra solutions appeared too. It would be interesting to find out what additional solutions will appear in our case. It is also worth investigating whether the warped anti-de Sitter solutions survive higher-order contributions~\cite{Moutsopoulos:2016bum}.

\subsection*{Acknowledgments}

We would like to thank Mehmet Ozkan for useful discussions. GM is fully and NSD is partially supported by the Scientific and Technological Research Council of
Turkey (T\"UB\.ITAK) project 113F034.

\appendix
\section{Proof of Null Supersymmetric Metric Form}\label{app:gib}
We sketch the proof for $D\neq0$ in \eqref{eq:gnullDnotzero}. Assume a null Killing vector $K$ for which \eqref{eq:beltrami} holds. Choose coordinates $(v,x^1,x^2)$ so that $K=\partial_v$,
\begin{equation}
g= h_{ij}(x)\grad x^i \grad x^j + A_i(x) \grad x^i \grad v~,
\end{equation}
where $h_{ij}$ and $A_i$ do not depend on $v$. If $D\neq 0$ then $\grad A \neq 0$ and we may choose coordinates $x^i\mapsto (u,\rho)$ so that
\begin{equation}
A = 2 e^{2\rho} \grad u~.
\end{equation}
The freedom
\begin{equation}
v \mapsto v + g(u,\rho)
\end{equation}
allows us to fix $h_{u\rho}=0$. We choose the sign of orientation so that $\ast \grad u = + h_{\rho\rho}\grad r \wedge \grad u$. With $A=K$ and
\begin{align}
\grad K &= 2 e^{2\rho} \grad \rho \wedge \grad u \\
\ast K &= \sqrt{h_{\rho\rho}}e^{2\rho} \grad \rho \wedge \grad u~,
\end{align} 
we equate the two sides of \eqref{eq:beltrami} in order to derive 
\begin{equation}
h_{\rho\rho}=\frac{1}{4D^2}~.
\end{equation}
We note that the sign of $D$ is correlated to the sign of orientation. That is, a null Killing vector for which \eqref{eq:beltrami} holds does not only 
fix the metric up to the undetermined $D$ and $h$, but it also fixes the sign of orientation. The metric is as in \eqref{eq:gnullDnotzero} in terms of an 
undetermined function $h_{uu}=h(u,\rho)$ and if  $\ast \grad u = + h_{\rho\rho}\grad r \wedge \grad u$ (respectively $- h_{\rho\rho}\grad r \wedge \grad u$) then $D<0$ 
(respectively $D>0$). 

\section{Killing Spinor Equation}
\label{app:ks}
Here we write the Killing spinor equations explicitly in the coordinate frame, first for the null supersymmetric background and then for the timelike supersymmetric background. The Killing spinor equation acting on a spinor $\epsilon'$ in the null background \eqref{eq:gnullDnotzero} is expanded into
\begin{align}
\left(\partial_\rho +\frac{1}{4} \frac{\partial_u g_{\rho\rho}}{\sqrt{g_{\rho\rho}}} \gamma^{\underline{\rho}+} - \frac{1}{2} \gamma^{+-} - D{\sqrt{g_{\rho\rho}}}\gamma_{\underline{\rho}} \right)\epsilon' &= 0\\
\left(\partial_u + \frac{1}{4 \sqrt{g_{\rho\rho}}} \left( h-\partial_\rho h\right) \gamma^{\underline{\rho}+} - \frac{1}{2\sqrt{g_{\rho\rho}}} \gamma^{\underline{\rho}-} - D \left( \gamma_+ + \frac{h}{2} \gamma_{-}\right)\right)\epsilon' &= 0\\
\left(\partial_v - \frac{e^{2\rho}}{2\sqrt{g_{\rho\rho}}}\gamma^{\underline{\rho}+} - D e^{2\rho} \gamma_- \right)\epsilon' &= 0~,
\end{align}
where $g_{\rho\rho}=1/(4D^2)$ is not necessarily constant. 
On the other hand, the Killing spinor equation acting on a spinor $\epsilon'$ in the timelike supersymmetric background \eqref{eq:KKform} is expanded into
\begin{align}
\partial_t \epsilon' &= -  \partial_t c  \, \gamma_{\underline{0}} \epsilon'
- i D \left( 1+ i \gamma_{\underline{0}}\right)\epsilon' - \frac{i}{2}(\partial_if )e^{-\sigma} \gamma^{\underline{i}}\left(1+i\gamma_{\underline{0}}  \right)\epsilon' 
\label{eq:kstimederivative}\\
0&=\partial_{\underline{i}} \epsilon' + i (\partial_{\underline{i}}c)\epsilon'
+ \frac{1}{2} e^{-\sigma}(\partial_i \ln f) i \gamma_{\underline{0}} \\ \notag &\quad+ \left( \frac{1}{4} f\rho \gamma_{\underline{i}}+ \frac{i}{2} e^{-\sigma}\epsilon_{ij}\partial_j\sigma- D\gamma_{\underline{i}}   \right)\left( 1+ i \gamma_{\underline{0}}\right)\epsilon'
\end{align}
Note how the gauge \eqref{eq:temporal} implies that the Killing spinor $\epsilon$, for which $\gamma_{\underline{0}}\epsilon=i\epsilon$, is time independent and the solution is indeed provided by \eqref{eq:kssol}.

\newpage



\begin{thebibliography}{10}
\bibitem{Achucarro:1987vz}
A.~Achucarro and P.~K. Townsend, ``{A Chern-Simons Action for Three-Dimensional
  anti-De Sitter Supergravity Theories},''
\href{http://dx.doi.org/10.1016/0370-2693(86)90140-1}{{\em Phys. Lett.} {\bf
  B180} (1986)  89}.

\bibitem{Uematsu:1984zy}
T.~Uematsu, ``{Structure of $N=1$ Conformal and Poincare Supergravity in
  (1+1)-dimensions and (2+1)-dimensions},''
\href{http://dx.doi.org/10.1007/BF01571396}{{\em Z. Phys.} {\bf C29} (1985)
  143}.

\bibitem{Rocek:1985bk}
M.~Rocek and P.~van Nieuwenhuizen, ``{N $>$= 2 Supersymmetric Chern-Simons
  Terms as D = 3 Extended Conformal Supergravity},''
\href{http://dx.doi.org/10.1088/0264-9381/3/1/007}{{\em Class. Quant. Grav.}
  {\bf 3} (1986)  43}.

\bibitem{Nishino:1991sr}
H.~Nishino and S.~J. Gates, Jr., ``{Chern-Simons theories with supersymmetries
  in three-dimensions},''
\href{http://dx.doi.org/10.1142/S0217751X93001363}{{\em Int. J. Mod. Phys.}
  {\bf A8} (1993)  3371--3422}.

\bibitem{Howe:1995zm}
P.~S. Howe, J.~M. Izquierdo, G.~Papadopoulos, and P.~K. Townsend, ``{New
  supergravities with central charges and Killing spinors in
  (2+1)-dimensions},''
  \href{http://dx.doi.org/10.1016/0550-3213(96)00091-0}{{\em Nucl. Phys.} {\bf
  B467} (1996)  183--214},
\href{http://arxiv.org/abs/hep-th/9505032}{{\tt arXiv:hep-th/9505032}}.

\bibitem{Cecotti:2010dg}
E.~Bergshoeff, S.~Cecotti, H.~Samtleben, and E.~Sezgin, ``{Superconformal Sigma
  Models in Three Dimensions},''
  \href{http://dx.doi.org/10.1016/j.nuclphysb.2010.04.023}{{\em Nucl. Phys.}
  {\bf B838} (2010)  266--297},
\href{http://arxiv.org/abs/1002.4411}{{\tt arXiv:1002.4411}}.

\bibitem{Bergshoeff:2010mf}
E.~A. Bergshoeff, O.~Hohm, J.~Rosseel, E.~Sezgin, and P.~K. Townsend, ``{More
  on Massive 3D Supergravity},''
  \href{http://dx.doi.org/10.1088/0264-9381/28/1/015002}{{\em Class. Quant.
  Grav.} {\bf 28} (2011)  015002},
\href{http://arxiv.org/abs/1005.3952}{{\tt arXiv:1005.3952}}.

\bibitem{alkac_massive_2015}
G.~Alkac, L.~Basanisi, E.~A. Bergshoeff, M.~Ozkan, and E.~Sezgin,
{``Massive {N}=2
  {Supergravity} in {Three} {Dimensions},''}  \href{http://dx.doi.org/10.1007/JHEP02(2015)125}{{\em JHEP}
 {\bf 1502} (2015) 125}, \href{http://arxiv.org/abs/1412.3118}{{\tt arXiv:1412.3118}}.

\bibitem{Kuzenko:2011xg}
S.~M. Kuzenko, U.~Lindstrom, and G.~Tartaglino-Mazzucchelli, ``{Off-shell
  supergravity-matter couplings in three dimensions},''
  \href{http://dx.doi.org/10.1007/JHEP03(2011)120}{{\em JHEP} {\bf 1103} (2011) 120},
\href{http://arxiv.org/abs/1101.4013}{{\tt arXiv:1101.4013}}.

\bibitem{kuzenko_three-dimensional_2011}
S.~M. Kuzenko and G.~Tartaglino-Mazzucchelli,
  {``Three-dimensional {N}=2
  ({AdS}) supergravity and associated supercurrents,''}\href{http://dx.doi.org/10.1007/JHEP12(2011)052}{{\em JHEP} {\bf 1112} (2011) 052},
  \href{http://arxiv.org/abs/1109.0496}{{\tt arXiv:1109.0496}}.

\bibitem{kuzenko_three-dimensional_2014}
S.~M. Kuzenko, U.~Lindstrom, M.~Rocek, I.~Sachs, and
  G.~Tartaglino-Mazzucchelli,
  {``Three-dimensional {N}=2
  supergravity theories: {From} superspace to components,''}\href{http://dx.doi.org/10.1103/PhysRevD.89.085028}{{\em Phys.\ Rev.\ D } {\bf 89} (2014) 8,  085028  }, \href{http://arxiv.org/abs/1312.4267}{{\tt arXiv:1312.4267}}.

\bibitem{Kuzenko:2015jda}
S.~M. Kuzenko, J.~Novak, and G.~Tartaglino-Mazzucchelli, ``{Higher derivative
  couplings and massive supergravity in three dimensions},''
  \href{http://dx.doi.org/10.1007/JHEP09(2015)081}{{\em JHEP} {\bf 09} (2015)
  081},
\href{http://arxiv.org/abs/1506.09063}{{\tt arXiv:1506.09063}}.

\bibitem{izquierdo_supersymmetric_1995}
J.~M. Izquierdo and P.~K. Townsend,
{``Supersymmetric
  spacetimes in 2+1 {adS}-supergravity models,''}
  \href{http://dx.doi.org/10.1088/0264-9381/12/4/003}{{\em Class. Quant. Grav.}  {\bf 12} (1995) 895},
  \href{http://arxiv.org/abs/gr-qc/9501018}{{\tt arXiv: gr-qc/9501018}}.

\bibitem{Deger:1999st}
N.~S. Deger, A.~Kaya, E.~Sezgin, and P.~Sundell, ``{Matter coupled AdS(3)
  supergravities and their black strings},''
  \href{http://dx.doi.org/10.1016/S0550-3213(99)00734-8}{{\em Nucl. Phys.} {\bf
  B573} (2000)  275--290},
\href{http://arxiv.org/abs/hep-th/9908089}{{\tt arXiv:hep-th/9908089}}.

\bibitem{AbouZeid:2001tu}
M.~Abou-Zeid and H.~Samtleben, ``{Chern-Simons vortices in supergravity},''
  \href{http://dx.doi.org/10.1103/PhysRevD.65.085016}{{\em Phys. Rev.} {\bf
  D65} (2002)  085016},
\href{http://arxiv.org/abs/hep-th/0112035}{{\tt arXiv:hep-th/0112035}}.

\bibitem{tod_all_1983}
K.~P. Tod, ``All {Metrics} {Admitting} {Supercovariantly} {Constant}
  {Spinors},'' \href{http://dx.doi.org/10.1016/0370-2693(83)90797-9}{{\em Phys.
  Lett. B} {\bf 121} (1983)  241--244}.

\bibitem{Andringa:2009yc}
R.~Andringa, E.~A. Bergshoeff, M.~de~Roo, O.~Hohm, E.~Sezgin, and P.~K.
  Townsend, ``{Massive 3D Supergravity},''
  \href{http://dx.doi.org/10.1088/0264-9381/27/2/025010}{{\em Class. Quant.
  Grav.} {\bf 27} (2010)  025010},
\href{http://arxiv.org/abs/0907.4658}{{\tt arXiv:0907.4658}}.

\bibitem{Deser:1982sw}
S.~Deser and J.~H. Kay, ``{Topologically Massive Supergravity},''
\href{http://dx.doi.org/10.1016/0370-2693(83)90631-7}{{\em Phys. Lett.} {\bf
  B120} (1983)  97--100}.

\bibitem{gibbons_general_2008}
G.~W. Gibbons, C.~N. Pope, and E.~Sezgin,{``The {General}
  {Supersymmetric} {Solution} of {Topologically} {Massive}
  {Supergravity},''}
  \href{http://dx.doi.org/10.1088/0264-9381/25/20/205005}{{\em Class.\ Quant.\ Grav.}  {\bf 25} (2008) 205005}, \href{http://arxiv.org/abs/0807.2613}{{\tt arXiv:0807.2613}}.

\bibitem{deger_supersymmetric_2013}
N.~S. Deger, A.~Kaya, H.~Samtleben, and E.~Sezgin, ``Supersymmetric {Warped}
  {AdS} in {Extended} {Topologically} {Massive} {Supergravity},'' \href{http://www.sciencedirect.com/science/article/pii/S0550321314001205}{{\em Nucl.\ Phys.\ B }{\bf 884} (2014) 106},
  \href{http://arxiv.org/abs/1311.4583}{{\tt arXiv: 1311.4583}}.

\bibitem{alkac_supersymmetric_2015}
G.~Alkac, L.~Basanisi, E.~A. Bergshoeff, D.~O. Devecioglu, and M.~Ozkan, ``Supersymmetric
  {Backgrounds} and {Black} {Holes} in  {N}=(1,1)
  {Cosmological} {New} {Massive} {Supergravity},''
 \href{http://dx.doi.org/10.1007/JHEP10(2015)141}{{\em JHEP} {\bf 1510} (2015) 141}, \href{http://arxiv.org/abs/1507.06928}{{\tt arXiv:1507.06928}}.

\bibitem{knodel_rigid_2015}
G.~Knodel, P.~Lisbao, and J.~T. Liu, ``Rigid {Supersymmetric} {Backgrounds} of
  3-dimensional {Newton}-{Cartan} {Supergravity},'', \href{http://arxiv.org/abs/1512.04961}{{\tt arXiv:1512.04961}}.

\bibitem{anninos_warped_2009}
D.~Anninos, W.~Li, M.~Padi, W.~Song, and A.~Strominger,{``Warped {AdS}\_3
  {Black} {Holes},''}
  \href{http://dx.doi.org/10.1088/1126-6708/2009/03/130}{{\em JHEP} {\bf 0903} (2009) 130}, \href{http://arxiv.org/abs/0807.3040}{{\tt arXiv: 0807.3040}}.

\bibitem{baum_twistor_2000}
H.~Baum, {``Twistor and {Killing} spinors
  in {Lorentzian} geometry,''} in \href{http://dx.doi.org/2-85629-094-9}{{\em Seminaires et {Congres} 4}, vol.~2000,
  pp.~35--52. 2000}.

\bibitem{lichnerowicz_spin_1987}
A.~Lichnerowicz,{``Spin manifolds,
  killing spinors and universality of the {Hijazi} inequality,''} \href{http://dx.doi.org/10.1007/BF00401162}{{\em Lett. Math.
  Phys.} {\bf 13} (1987)  331--344}.

\bibitem{rademacher_generalized_1991}
H.-B. Rademacher, ``Generalized killing spinors with imaginary killing function
  and conformal killing fields,'' in {\em Global {Differential} {Geometry} and
  {Global} {Analysis}}, D.~Ferus, U.~Pinkall, U.~Simon, and B.~Wegner, eds.,
  no.~1481 in Lecture {Notes} in {Mathematics}, pp.~192--198.
Springer Berlin Heidelberg, 1991.
\href{http://link.springer.com/chapter/10.1007/BFb0083642}{DOI: 10.1007/BFb0083642}.

\bibitem{mitsuka_no_2012}
Y.~Mitsuka and G.~Moutsopoulos,{``No more {CKY}
  two-forms in the {NHEK},''}
  \href{http://dx.doi.org/10.1088/0264-9381/29/4/045004}{{\em Class. Quant. Grav.} {\bf 29} (2012)  045004}, \href{http://arxiv.org/abs/1110.3872}{{\tt  arXiv: 1110.3872}}.

\bibitem{moutsopoulos_homogeneous_2013}
G.~Moutsopoulos,{``Homogeneous
  anisotropic solutions of topologically massive gravity with cosmological
  constant and their homogeneous deformations,''}
  \href{http://dx.doi.org/10.1088/0264-9381/30/12/125014}{{\em  Class. Quant. Grav.} {\bf 30} (2013)  125014},
  \href{http://arxiv.org/abs/1211.2581}{{\tt arXiv:1211.2581}}

\bibitem{misner_taub-nut_1965}
C.~W. Misner, ``Taub-nut space as a counterexample to almost everything,'' 1967, Relativity 
Theory and Astrophysics.~Vol.1: Relativity and Cosmology, 1, 160 

\bibitem{moutsopoulos_nut_2010}
G.~Moutsopoulos,{``The
  {NUT} in the {N}=2 {Superalgebra},''} \href{http://dx.doi.org/10.1088/0264-9381/27/3/035008}{{\em Class. Quantum Grav.} {\bf 27}
  (2010)  035008}, \href{http://arxiv.org/abs/0908.0121}{\tt arXiv:0908.0121}.

\bibitem{anninos_curious_2010}
D.~Anninos, G.~Compère, S.~de~Buyl, S.~Detournay, and M.~Guica,{``The {Curious} {Case} of
  {Null} {Warped} {Space},''}
  \href{http://dx.doi.org/10.1007/JHEP11(2010)119}{{\em JHEP} {\bf 1011} (2010) 119 }, \href{http://arxiv.org/abs/1005.4072}{\tt arXiv:1005.4072}.

\bibitem{jugeau_accelerating_2011}
F.~Jugeau, G.~Moutsopoulos, and P.~Ritter,{``From accelerating
  and {Poincar}{\textbackslash}'e coordinates to black holes in spacelike
  warped {AdS}3, and back,''}
  \href{http://dx.doi.org/10.1088/0264-9381/28/3/035001}{{\em  Class. Quant. Grav.} {\bf 28}
  (2011)  035001}, \href{http://arxiv.org/abs/1007.1961}{\tt arXiv:1007.1961}.

\bibitem{spradlin_vacuum_1999}
M.~Spradlin and A.~Strominger,{``Vacuum {States} for
  {AdS}2 {Black} {Holes},''}
  \href{http://dx.doi.org/10.1088/1126-6708/1999/11/021}{{\em JHEP} {\bf 9911} (1999) 021}, \href{http://arxiv.org/abs/hep-th/9904143}{\tt arXiv:
  hep-th/9904143}.

\bibitem{hawking_soft_2016}
S.~W. Hawking, M.~J. Perry, and A.~Strominger, ``Soft {Hair} on {Black}
  {Holes},''
  \href{http://arxiv.org/abs/1601.00921}{\tt arXiv:1601.00921}.

\bibitem{coussaert_supersymmetry_1994}
O.~Coussaert and M.~Henneaux, ``Supersymmetry of the (2+1) black holes,''
  \href{http://dx.doi.org/10.1103/PhysRevLett.72.183}{{\em Phys.Rev.Lett.} {\bf
  72} (1994)  183--186}.

\bibitem{coussaert_selfdual_1994}
O.~Coussaert and M.~Henneaux, ``Selfdual solutions of (2+1) {Einstein} gravity
  with a negative cosmological constant,''
  In Teitelboim, C. (ed.): The black hole 25-39
   \href{http://arxiv.org/hep-th/9407181}{{\tt arxiv:hep-th/9407181}}.

\bibitem{Moutsopoulos:2016bum}
  G.~Moutsopoulos,
  ``Warped anti-de Sitter in 3d (2,0) Supergravity,''
  \href{http://arxiv.org/abs/1602.08733}{{\tt arXiv:1602.08733}}.
  
\end{thebibliography}
\end{document}